\documentclass[aps,preprint,showpacs,showkeys]{revtex4}
\usepackage{amssymb,amsmath,epsf,graphicx}
\usepackage[ansinew]{inputenc}

\begin{document}
\title{A systematic study of non-ideal contacts in integer quantum Hall systems}

\author{Christoph Uiberacker}
\email{christoph.uiberacker@unileoben.ac.at}

\author{Christian Stecher}

\author{Josef Oswald}

\affiliation{Institute of Physics, University of Leoben, 
Franz-Josef-Strasse 18A, 8700 Leoben, Austria}

\begin{abstract}
In the present article we investigate the influence of the contact region on the distribution of the chemical potential in integer quantum Hall samples, as well as the longitudinal and Hall resistance as a function of the magnetic field. First we use a standard quantum Hall sample geometry and analyse the influence of the length of the leads where current enters/leaves the sample and the ratio of the contact width to the width of these leads. Furthermore we investigate potential barriers in the current injecting leads and the measurement arms in order to simulate non-ideal contacts. Second we simulate nonlocal quantum Hall samples with applied gating voltage at the metallic contacts. For such samples it has been found experimentally that both the longitudinal and Hall resistance as a function of the magnetic field can change significantly. Using the nonequilibrium network model we are able to reproduce most qualitative features of the experiments.
\end{abstract}
\maketitle

\newpage 

\section{Introduction}
Nowadays the steps in the transversal (Hall) resistance at the inverse of integer multiples of $e^2/h$ are well understood in terms of one-dimensional edge channels \cite{datta95}. The number of such channels decreases with increasing magnetic field due to increasing energy spacing between the Landau levels.

Despite this clear physical picture it was found experimentally that width and magnetic field values of the transition region between plateaus can change in case of nonideal contacts, e.g.~when applying gating to the contacts. 
Since a few years experiments using scanning probe techniques (see Ahlswede et.~al.~\cite{ahlswede}) were performed to visualize the distribution of the nonequilibrium potential near non-ideal contacts. On the other hand also the distribution of current in quantum Hall samples was investigated \cite{dominguez89} in order to optimize the contact geometry. A significant number of experiments (see \cite{alphenaar90,mueller90,dahlem}) found a strong indication that the contacts do not behave as ideal ones, meaning that not all edge channels seem to arrive at the reservoir of the metallic contacts of the sample. In addition imperfect equilibration among edge channels can also lead to deviations in the measured resistance values.  Furthermore, deviations of the $R_{xx}$ peaks from the expected shapes were observed for non-ideal contacts by Dahlem et.~al.~\cite{dahlem}, who proposed that the deviation might result from the crystalline orientation of the edge of the contacts. The dependence of the contact resistance on the crystal orientation has also been subject to other experiments \cite{goektas}. They embedded Au/Ge/Ni contacts in various $Al_{0.67}Ga_{0.33}As/GaAs$ heterostructures and found that the contact resistance varied with length and orientation of the interface line of the contact.

The experimental findings show nontrivial results for the potential distribution and resistances as a function of details of the sample geometry. In order to simulate such samples we use the nonequilibrium network model (NNM) (Ref.~\cite{oswald98} and \cite{oswald06}) because the actual geometry of the sample can be taken into account. In addition the NNM has already proven successful to simulate even complicated sample geometries like anti-Hall bars \cite{antiHallbar}.
 
Regarding theoretical investigations of nonideal contacts in simplified models, van Wees et.~al.~discuss influences of non-ideal contacts in terms of a nonzero reflection probability of electrons at the contact, which leads to waves traveling in the system \cite{vanWees90}. This leads to potential resonances and a fine-structure of the conductance between plateaus. We did not consider such effects here and assumed zero reflection at all our contacts. Note in case of slowly varying electrostatic potential from the system into the contact the reflection would be suppressed \cite{glazman}.

The article is structured as follows: In Sec.~\ref{s_theory} we present the setup and underlying theory of the NNM. We investigate the influence of the geometric shape of the contact and the current-injecting leads as well as potential barriers within leads and measurement contacts in Sec.~\ref{s_standardgeom} of the results. Simulations of gated samples are then presented in section \ref{s_gatings}. Finally we summarize and conclude in Sec.~\ref{s_conclusion}.

\section{Theory}
\label{s_theory} 
 
The exact Hamiltonian of an integer quantum Hall sample is not easily formulated 
because the disorder potential (long ranged), which defines the underlying electrical potential besides  the atomic (microscopic) environment and the confining potential (due to interfaces), depends on details of the exact configuration within the sample. This lead to a variety of more or less simplifying models.
The most prominent of this type of models is the Chalker-Coddington model \cite{ccm}, which is able to predict delocalized states but cannot describe the nonequilibrium steady state. However it seems to lead to the correct exponent of the correlation length \cite{hohl03,cain01,cain04} from which we conclude that the topology of the setup of the network is sensible even in the nonequilibrium regime.

Due to the strong magnetic fields Landau levels develop with localization within the magnetic length $l_B=\sqrt{\hbar/eB}$ given by the cyclotron motion, $e$ and $B$ denote elementary charge and magnetic induction. Considering the very different length scales involved, electrons are confined to move semi-classically along equipotential lines of the disorder potential \cite{localizedstates}. 
Regarding this approximation transport occurs via tunneling constrictions, that is, saddle points in the underlying potential landscape. The elastic tunneling transition probabilities across such saddles have been calculated quantum mechanically for equilibrium \cite{fertig87,buettiker90}.

In order to be able to discuss special wiring configurations we have to define the notion of longitudinal and transversal components with the help of a curvilinear coordinate system, determined by the local orientation of the contour of the chemical potential in the plateau regime. In this way the local longitudinal direction is defined within the plateau state via the local tangent to the equipotential line. Clearly no distinction between longitudinal and transversal resistance can be made in the transition region. If then two measurement contacts can be connected within the plateau region along an equipotential line, we measure a longitudinal resistance. In case of intersections a transversal (Hall) component is measured. Note however, the NNM calculates within a fixed coordinate system labeled by $(x,y)$. As required the resistances do not depend on the choice of these coordinates.
  
The physical content of the NNM can be understood in terms of local equilibrium (see e.g.~Ref.~\cite{zubarev} for a formulation in continuous space variables) when reformulating it for the network of semiclassical wavefunctions. In this way we attribute unique thermodynamical quantities such as the chemical potential to each single wavefunction, given by a trajectory along the contour connecting two saddle points.

Oh and Gerhardts \cite{gerhardts} use a very related concept of local equilibrium, however in a continuum description, together with the Thomas-Fermi or Hartree approach to the self-consistent calculation of charges. They use a simple model geometry and assume translational invariance in longitudinal direction which results by continuity in a current independent of the transversal direction. They also include the velocity of particles when averaging over states to obtain the charge density. Current injection and the important influence of contacts are however not discussed. 

\subsection{Setup of the NNM}

We start by replacing the underlying potential landscape, consisting of the long-ranged random potential due to impurities and doping atoms and the confining potential due to the surface of the sample, by a regular grid of nodes. This is fairly general because it is possible to model trajectories of arbitrary shape by states with appropriate energies at the saddles. On this background we replace the random potential by an effective oscillating potential $V(x,y)=\tilde{V}[\cos(\omega y)-\cos(\omega x)]$ with period $L:=2\pi/\omega$ and amplitude $\tilde{V}$ obtained as averages of the random potential. The potential distribution is then only topologically changed (e.g.~deformations of contours) but edge channels and backscattering remain. 

\begin{figure}
 \begin{center}
   		\includegraphics[scale=0.7]{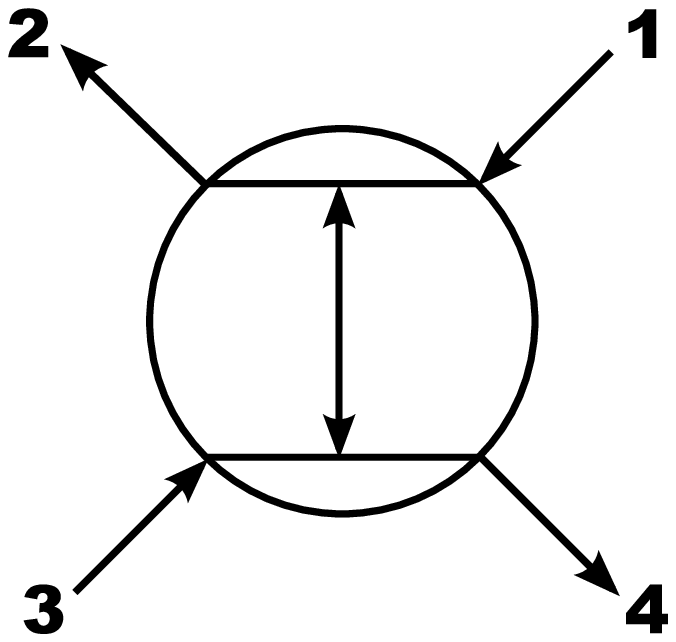}
   \caption{}
   \label{f_node}
 \end{center}
\end{figure}

Due to the topology of the saddle points each node connects to four chemical potentials and is visualized as a circle with links labeled by $u_1,\dots,u_4$ counter-clockwise with $u_1$ denoting the upper, right corner (see Fig.~\ref{f_node}). Consistency demands that only two independent differences of potentials exist, which represent the two components of the local electric field. Opposite differences are therefore equal \cite{oswald06}. We then define the ratio of longitudinal to transversal field component, given by
\begin{equation}
	P := \frac{E_x}{E_y}=\frac{u_1-u_2}{u_1-u_4}=\frac{u_4-u_3}{u_2-u_3} \quad .
\end{equation}
In this way we construct a ''transfer'' equation for the chemical potentials
\begin{equation}
	\left[\begin{array}{c} u_2 \\ u_3 \end{array}\right] =
		 \left[\begin{array}{cc} 1-P & P \\ -P & 1+P \end{array}\right]
		 \left[\begin{array}{c} u_1 \\ u_4 \end{array}\right] \quad .
\end{equation}
The chemical potential distribution can be calculated as a boundary value problem once the values for $P$ at each node are given. External contacts supplying current are modeled by saddles with a pair of trajectories that point into/out of the sample ($E_x=0$, compare the Landauer-Büttiker picture \cite{datta95}) and one of these two trajectories is kept fixed at the chemical potential of the supply contact.

The random potential is mapped to the NNM by setting up the grid in such a way that in every second column the nodes are rotated clockwise by an angle of 90° and  $P'=1/P$ is used for $P$. This reflects the topology of saddle points of the underlying oscillating effective potential $V(x,y)$. As a result loops can be formed in isolated valleys or around peaks. Such loops have actually been observed experimentally \cite{CarrierLoops}. 

We use the Landau levels and a self-consistent Thomas-Fermi approximation in order to obtain the charge density at zero temperature. Finite temperatures increase the transition regions as was analysed in a previous paper \cite{oswald98_2}. The screening of the electrical bare potential is then obtained from the charge density simply by multiplication with a given constant factor of $C=50mV/(10^{11}cm^{-2})$ (units $e=1$). Furthermore we use a constant broadening of $0.5meV$ to mimic Landau bands generated by the disorder potential. 
Concerning spin splitting a $g$-factor of $g=4$ was used for all calculations \cite{gfactor}.

We neglect effects due to feedback of the chemical potential onto the charge distribution and hence on the values of $P$. Note this amounts to the assumption that states across saddles describe a linear interpolation between states of two chemical potentials, that is, the system traverses from equilibrium to the nonequilibrium steady state adiabatically.

In order to calculate the $P$-values we adopt the high-field approximation of localized electron states in form of contour lines of the potential, also at saddle points. In this way we assume a purely off-diagonal conductance tensor within the sample with a contribution of $e^2/h$ from each Landau level. Then we can immediately write $P=I_y/I_x$ as a ratio of transversal to longitudinal component of the current.

According to the edge channel picture we call $T$ the probability of transmission in longitudinal direction, therefore we get $I_y\propto R$ and $I_x\propto T=1-R$. This can be used to calculate $P$ from details of the Landau levels.
Compare the work of Streda et.~al.~\cite{streda87} in this respect, who describe the whole sample as such a single node. The current of the sample is finally calculated by taking the sum of all transversal potential differences in the contact nodes and multiply by $e^2/h$. Equality of total input and output current determines the potential distribution uniquely.  

The present algorithm has the advantage that only the electric field has to be calculated. The only detail used from the Landau levels is the energy spacing of levels and the number of levels below $E_F$. 
Similar to other approaches \cite{gerhardts,nachtwei} we are able to calculate longitudinal and transversal resistance by identifying the current with the macroscopic current direction. The local dissipative component could be found with a formulation of tunneling and electron statistics in nonequilibrium together with a principle to describe the nonequilibrium steady state, such as, e.g., minimum entropy production, which would however make calculations much more demanding. Such investigations are under way and will be published elsewhere. 
 
In this spirit we use the expression of the tunneling transmission $R_{mn}$ between edge channels through the saddle in the presence of a magnetic field derived in Refs.~\cite{fertig87} and \cite{buettiker90} to obtain $P$ (Ref.~\cite{oswald06})
\begin{equation}
	P = \delta_{mn}\frac{R_{mn}}{1-R_{mn}}= 	
		\exp\left[-\frac{L^2B}{h\tilde{V}}\epsilon\right] \quad ,
\end{equation}
with $\epsilon:=E_F - V_S$ the difference of the Fermi energy to the saddle energy $V_S$ and $B$ denoting the magnetic field strength.

In the plateau region the loops in the bulk are isolated but in the transition between plateaus a fraction of electrons in the edge channels tunnels into the bulk, leading to a potential difference in the direction of the mainly longitudinal current and therefore dissipative transport.
Equilibration among edge channels is allowed by simulating tunneling between edge channels with a chosen constant decay length for the involved states.  
   
\section{Results}

\subsection{Standard sample geometry}
\label{s_standardgeom}

In this paper we present a systematic study of nonideal contacts. The sample is set up as a typical Hall bar with two current-injecting metallic contacts and four measuring contacts to obtain longitudinal and transversal resistances. In the course we investigate the distribution of the chemical potential across the sample when varying the ratio of contacted cross-section of the leads as well as the length of the leads. Furthermore we placed a barrier across the width of the leads in vicinity of the current-injecting contacts (contact nodes), formed by a potential energy ridge consisting of a plateau and quadratic tails, to simulate non-ideal matching between the states in the contact and the 2d electron gas. The curvature of the saddles was chosen as $a:=2h\tilde{V}/L^2=1$ in all calculations of this section.

Regarding the metallic contacts and the leads, the quantities relevant within the nonequilibrium steady state are the ratio $r_c:=W_C/W_L$ of the width of the contact $W_C$ to the width of the lead $W_L$ and the ratio $r_l:=L/W$, $0\le r_l<\infty$, of the lead length $L$ to its width $W$. We motivate this choice by noting that the potential distribution tends to adapt to the sample geometry and therefore simply scales on multiplying $x$ and $y$ with a common scale factor. 

We present an overview of transversal and longitudinal resistances against magnetic field in Fig.~\ref{f_RLeads} by comparing samples with different $r_l$. In Figs.~\ref{f_RContactdiff} and \ref{f_Rdiff} we show differences to illustrate small changes while changing $r_c$ and $r_l$. Introducing a barrier in the current injecting leads is discussed in Fig.~\ref{f_BarrierRdiff}. The relative error of Hall resistances for various calculated setups is demonstrated in Fig.~\ref{f_RrelErr}.
  
\subsubsection{Different metallic contacts}

We investigate two limits when applying the external potential, namely point contacts ($r_c\to 0$) on the one hand and second contacting the whole cross section of the lead ($r_c=1$). Intuitively for long enough leads we expect the potential to adapt to the shape of the lead and the geometrical shape of the contact should be irrelevant. This has indeed been found in experiments \cite{LeadLengthConvergence}, the rule of thumb to be sure that the measurement does not depend on the form of the contact is to use leads with $r_l\ge 4$. 

In Fig.~\ref{f_contacts} we compare the potential distribution for samples with $r_c\to 0$ and $r_c=1$ in the plateau and transition region, using $r_l=1.5$ for both samples. It turns out that the potentials are very similar in the whole sample except near the contacts. However we note that in contrast to the rule of thumb the length-scale around the contact where the potential contours have different slopes is only a fraction of the width of the lead. In the plateau region this length is even smaller than in the transition region, because $P\approx 0$ near the boundary forces the potential contours parallel to the longitudinal direction. 

In Fig.~\ref{f_RContactdiff} we show the difference between resistances for point contacts and the respective values for $r_c=1$. We obtain very different magnitude of differences depending on $r_l$ and on the magnetic field $B$. Apparently the differences are very small for $r_l=1.5$ up to a field of about $10T$. On the other hand, for $r_l=0$ the differences are much larger and one can clearly see $\Delta R_{xy}$ tracing the structure of peaks in $R_{xx}$. Furthermore it is interesting to note that independent of $r_l$ and $B$ the difference of $R_{xx}$ is significantly smaller than for $R_{xy}$. We conclude that a difference between point- and line-contacts vanishes quickly with increasing $r_l$ and $R_{xy}$ is stronger influenced. 

\subsubsection{Length of the leads}

We calculated five sample geometries, one with practically no leads (limit $r_l\to 0$) , one with small leads ($r_l = 0.5$), one with medium length ($r_l=1.5$) and two with long leads ($r_l=3$, $r_l=5$). Chemical potential distributions at selected magnetic fields are shown in Fig.~\ref{f_lengthLead}. To our surprise even for $r_l<<4$ (at least down to $r_l=1$) the field contours adapt from the contacts to the parallel configuration in the bulk within the plateau region before leaving the leads, against the rule of thumb. Furthermore it seems that actually the lengthscale for this adjustment depends very much on the length of the leads itself. The adjustment is the faster the shorter the leads, again pointing to a scaling behaviour of the chemical potential. We can then understand this behaviour by noting that the change in geometry when leaving the lead, in our case the widening of the sample by the measurement contacts, forces the potential to adapt to the core of the sample.  

The calculated resistance differences from changing $r_l$ restore the rule of thumb, meaning that resistance values converge nonlinearly with increasing $r_l$, with $r_l=3$ already close to the limit.  
In Fig.~\ref{f_RLeads} we find the typical function of transversal and longitudinal resistances with the magnetic field. At low field values the peaks of the longitudinal resistance clearly show that spin levels are not completely resolved as the 2 pairs around $2.5T$ and $3-4T$ are not yet split.
We then show the difference of transversal and longitudinal resistance of various values of $r_l$ to the corresponding resistances for long leads, $r_l=5$ in Fig.~\ref{f_Rdiff}. It is apparent that $R_{xx}$ decreases slightly with increasing $r_l$ while $R_{xy}$ increases at the same time. We also see a convergence with increasing $r_l$, that is the differences for $r_l=1.5$ are smaller than half the one from $r_l=0$ and only a few ohms at $r_l=3$. One further notes that differences of $R_{xx}$ and $R_{xy}$ have opposite sign and approximately equal magnitude. Making the plausible assumption that the current does hardly change with $r_l$, this points to the fact that the sum of transversal and longitudinal potential difference is approximately constant for various $r_l$. This can be explained qualitatively by the fact that all chemical potentials are evaluated by current conservation, because then the sum of potentials should stay constant if the current does not change. This also gives a hint, why $\Delta R_{xx}$ and $\Delta R_{xy}$ in Fig.~\ref{f_RContactdiff} are not symmetric around zero in case of altering $r_c$. This time we expect the current to change with the contact geometry.   

\subsubsection{Contact barriers}

All previous calculations have been made under the assumptions that the equilibrium electrical potential within the leads is fairly flat. This is somehow unrealistic because the interface between metal and 2DEG usually develops a barrier due to charge transfer processes, having different potential energy. We try to mimic realistic contacts by introducing such a barrier in front of the metallic contacts. The height of the bare barrier was chosen to be $50meV$ at its plateau, which is reduced by screening to a value in the range $0.5meV$ - $8.5meV$ depending on the Fermi energy. The distribution of the chemical potential is shown in Fig.~\ref{f_barrier} for low and high magnetic field in the plateau and transition region for the case of each current-injecting contact having a barrier. Furthermore we compare the scenario of having a barrier in only one of the leads for the transition region within the high field case. The most striking difference between low and high field is the larger angle of equipotential contours at high field relative to the longitudinal direction in the bulk for low magnetic field. This agrees with the increasing peak maxima of the longitudinal resistance when increasing the magnetic field. There are two aspects for understanding this behaviour. We note that $P=0$ (transport by edge channels only) holds for all network layers except the topmost one. Therefore the decreasing number of Landau levels decreases the total current and consequently increases the resistances because in these layers with $P=0$ each one contributes the same part to the total current. On the other hand with increasing number of layers with $P=0$ the potential distribution is forced stronger to a plateau-like potential distribution even if the topmost layer has $P>0$ in the transition region.  

We compare the influence of various configurations of barriers on the longitudinal resistance in Fig.~\ref{f_RBarriers}. Obviously considering barriers in the measurement contacts has a larger effect than non-ideal current-injecting leads alone. We note that the transition starts at slightly lower field for barriers in all arms and the transition region is enlarged.

In Fig.~\ref{f_BarrierRdiff} we plotted the difference in resistances for the left lead or both leads having a barrier and the corresponding values for the sample without barriers. Interestingly $R_{xx}$ decreases and $R_{xy}$ increases when having a barrier in the leads, similar to the behaviour of the resistances with increasing $r_l$. The difference between one and both leads with barrier is fairly small, especially for the transversal resistance. Both leads having a barrier produced a larger $|\Delta R_{xx}|$ than one barrier, as expected. In addition we compared calculations with barriers of bare height $100meV$ in the current injecting leads
with the analogous calculations for $50meV$ described above. The difference of resistances is significantly smaller than the comparison to the case of no barriers. This can be understood by the fact that the height after screening is not very different despite the ratio of 2 for the unscreened barriers. However we observe a much larger error for samples with barriers in the measurement contacts. This is to be expected because these barriers directly influence the potential differences measured. Apparently this has a larger effect than a change in total current due to barriers in the current-injecting leads.

\subsubsection{Dissipation}

We can learn valuable details about dissipation by inspecting the chemical potential distribution. We find in Fig.~\ref{f_contacts} that the equipotential contours always form loops that start from above a current-inducing contact and end below the {\it same} contact. Therefore, due to the setup that a finite current is flowing from one to the other contact it is necessary for this current to cross contours, meaning that the current flowing through the sample must have a dissipative component. 
Assuming homogeneous diagonal entries to the conductivity tensor at saddles across the sample (which should be an adequate approximation for a constant equilibrium potential energy across the sample) we find that the dissipative current should be inversely proportional to the gradient of the potential. 
We conclude that entropy production occurs quite locally at the lower end of the left current-injecting contact and at the upper end of the right current-injecting contact in the geometry of our samples. This is in accord with measurements \cite{kent91}. Especially for contacts with barriers we expect hardly any current crossing the barrier but rather it will follow potential contours around the barrier. This is always possible if the barrier is not too high, as in our simulated cases.
We note that flow along the equipotential contours agrees with the experience that current flows along the edges \cite{vKlitzing95}, whereas the Landauer-Büttiker picture would give maximal current in the bulk as there are the maximum number of channels up to the Fermi energy \cite{tsemekhman95}.

\subsection{Nonlocal configurations and gated samples}
\label{s_gatings}

After investigating systematic changes at the contacted leads where current is injected into the sample we proceed to simulate nonlocal geometries of recent experimental investigations of non-ideal contacts with gatings.

To set up the simulation we first generated two samples with the same geometry as used in the experiment. In the following the contacts are numbered by starting with 1 at the left boundary edge and increasing the numbers in the clockwise direction. The first sample (S1) had the external potential connected to contacts 1 and 2 and the voltage drop is measured between contacts 4 and 3. In this way we measure a longitudinal voltage difference because (4,3) can be connected along equipotential lines in the plateau region.  The second sample (S2) had the external potentials connected to contacts 2 and 4 and measures the voltage difference between contacts 1 and 3. This therefore amounts to a transversal voltage drop. 

We ran calculations of both sample geometries by sweeping the magnetic field up to $12T$. We used ideal samples without gating and ones where gating of some of the contacts has been applied in analogy to the experiment. The samples are illustrated in Fig.~\ref{f_sample1} and \ref{f_sample2}. 
In all calculations of this subsection the saddle curvature was chosen to be $a:=2h\tilde{V}/L^2=0.1$.

Before discussing the potential distributions, we want to point out a general problem in this context. In analogy to the experiment the bulk region of the sample carries electrical potential even in case it is insulating (plateau regime). This potential distribution is left over from the previous transition between plateaus, after which the bulk became decoupled from the edges due to insufficient equilibration time. 
Consequently the bulk potential has physical relevance only within the transition between plateaus. 
 
\subsubsection{Enhanced longitudinal resistance in sample 1}

By sweeping the field we found that the sample with gating on contacts 1 and 4 shows an extra $R_{xx}$ peak centered around $B=5.7T$ and a significantly enlarged peak in the interval $[8.5T,11T]$, that is at the same time much broader than the respective peak of the sample without gates. Furthermore it forms a plateau in the middle, however with a dip around $B=10.33T$. The plot is given in Fig.~\ref{f_Rxx}. It is apparent that besides the two intervals $[5.5T,6T]$ and $[8.5T,11T]$ the samples with and without gating show qualitatively the same resistance, even with slightly higher values for the ideal sample.  

To understand all these features we concentrate on the potential distribution across the sample, which is given in Fig.~\ref{f_PotDistL} for four magnetic fields within or in vicinity to the larger peak of the longitudinal resistance of the gated sample. At $8.00T$ we still have perfect edge channels and therefore $R_L$ is zero. At around $8.5T$ dissipation sets in and $R_L$ becomes nonzero. Concentrating on the distribution at $8.86T$ we find the reason in an equilibration of channels that occurs at contact 3. Around a magnetic field of $10.3T$ a dip of $R_L$ develops which can be attributed to equilibration among edge channels. Note the edge channel with the potential of contact 1 strongly equilibrates at the lower left corner with other edge channels which are at the potential of contact 2 and have not been equilibrated to the potential of contact 1 yet due to the gating. This mechanism results in less voltage drop between contacts 3 and 4 because the main drop occurs in the lower left corner. Finally around $11T$ the dissipation vanishes in the gated and ungated samples and we reach the next plateau region. Note similar equilibrations among edge channels is responsible for the non-ideal dissipation producing the peak starting at $8.86~T$ and also the peak centered around $5.7T$. 

\subsubsection{Shift of transition in transversal resistance in sample 2}
  
For this calculation we used sample 2 with different gating configurations. Figure~\ref{f_Rxy} shows the transversal resistance obtained from the potential applied via contacts 1 and 3. There are no qualitative differences between the ungated sample and gating configurations a) and b). However gating all contacts leads to the interesting phenomenon of shifting the last plateau transition to a lower value of $B$ by the remarkable values of about $2T$. 
Furthermore for gating configuration c) a peak develops around $5.75T$ due to changes in edge channel equilibration, where all other gating configurations are within the plateau. It seems that also this transition starts shifted to lower magnetic fields but then falls back to the plateau value. This can happen when the equilibration region among edge channels changes from outside the range between the measurement contacts to inside. 

As an intuitive explanation for the shift of the last plateau transition it comes to mind that due to gating at all contacts isolated edge channels can be generated, which couple to only one contact. Noting that the energy of the states qualitatively follows the underlying potential energy surface and the Fermi energy in equilibrium is to a large extend given by bulk properties we arrive at the conclusion that indeed the number of occupied states is decreased at the gating. The plots of the distribution of chemical potentials in Fig.~\ref{f_decoupledEC} indeed support this explanation. We find that at $8.3~T$ with and without gating the edge channels from contact 2 to contact 4 have the same potential, due to equilibration at contact 4 where the external potential is applied. On the other hand for $9.5~T$ the gating configuration c) clearly develops two edge channels between contacts 2 and 4 of different potentials, namely, one with the potential of contact 4 and the other of contact 2. This results in an isolated edge channel that cannot contribute to the current and effects the shift of the transition.

\section{Summary and Conclusion}
\label{s_conclusion} 

In the present paper we summarize a series of calculations of resistances and chemical potential distributions for integer quantum-Hall samples with the aim to investigate the influence of the shape and potential energy of nonideal contacts where the current enters/leaves the sample. In the first part of the paper we investigate the influence of the ratio of the contacted part of the cross-section of the leads, the length of the leads and potential barriers within the leads and measurement contacts. The contours of the chemical potential more or less adapt to the local geometry, for instance they ''flow'' around barriers. From potential contours we were able to identify the regions where dissipation will occur. 
Resistances strongly depend on the details of the sample used. We find relative errors that range up to a few percent in the transition region. We conclude that the Quantum Hall effect is robust against such changes only within the plateau regime, that is, whenever the Fermi energy lies between two Landau levels.   

In the second part we calculated gated quantum-Hall samples in nonlocal geometries that have recently been used in experiments of non-ideal contacts. Various different configurations of gating at the metallic contacts are considered in analogy to the experiment. We find that partial equilibration near a remote gated contact can lead to enhanced dissipation. On the other hand if all involved contacts are gated it might happen at sufficiently high magnetic fields that one Landau level is connected to only one of the contacts and as a consequence does not contribute to transport and therefore the corresponding plateau transition is shifted down in magnetic field.
We believe that our calculations give important information for experimentalists, especially regarding questions of metrology.

Recent experiments \cite{baumgartner07} used a scanned tip to apply local gating and investigated the change in resistance. In this way the underlying potential landscape can be locally changed. Using a network model that combines tunneling and superconducting states the current distribution and hot spots have been calculated \cite{dubi06}. It would be an interesting extension of the present work to directly simulate these resistance changes, which is possible with the NNM. Future work is planned on this subject.  

\begin{acknowledgments}                                                                
This work was sponsored by the Austrian Science Fund under the
Research Program No.~P 19353-N16. 
\end{acknowledgments}

\newpage

\setlength{\parindent}{0cm}
\setlength{\parskip}{0.5cm}

{\bf Figure captions}

Figure 1. Representation of tunneling at a single saddle point. The saddle is denoted by a circle and the lines crossing the saddle are trajectories of edge channels. The double-arrow denotes tunneling between the edge channels.

Figure 2. (color online)
Comparison of point contacts with contacts extending over the width of the current-injecting leads. Visualization of chemical potential distributions in the sample. The left column corresponds to the plateau region at $9T$ whereas for the right column shows the region of transition to the last plateau at a magnetic field of $11T$. Positive/negative values of the chemical potential are shown by full/broken contours. We plotted the values $0,\pm 4,\pm 8$ with the largest/smallest value drawn in bold. The value $0$ has a separate linestyle for easy distinction. 

Figure 3. (color online)
Difference of transversal and longitudinal resistances for point contacts ($r_c=0$) relative to values for line contacts ($r_c=1$), for $r_l=0$ and $r_l=1.5$, plotted against magnetic field.

\begin{figure}
 \begin{center}
   \begin{tabular}{cc}
   		\includegraphics[scale=0.7]{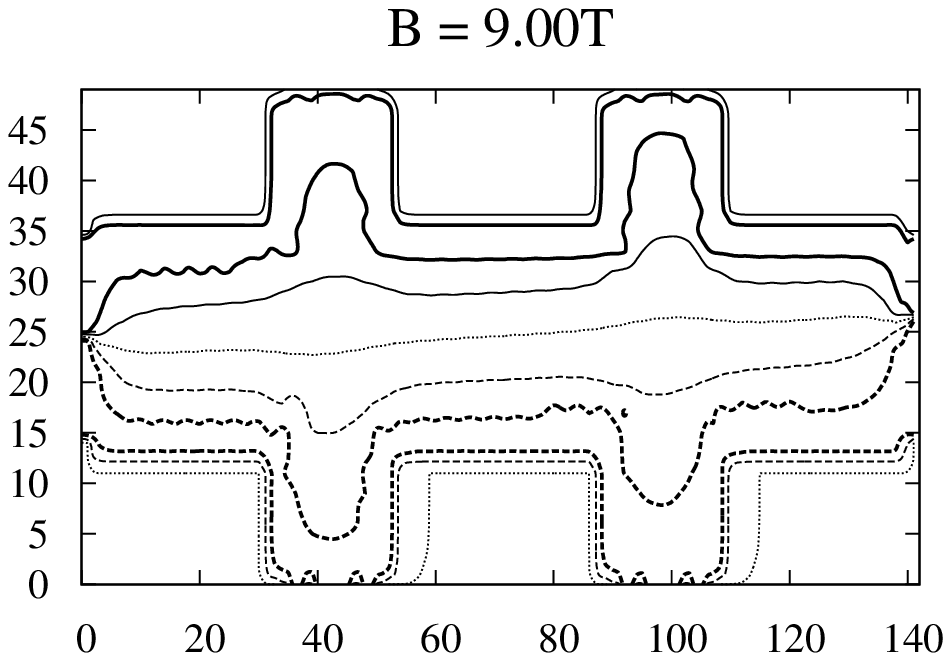} &
   			\includegraphics[scale=0.7]{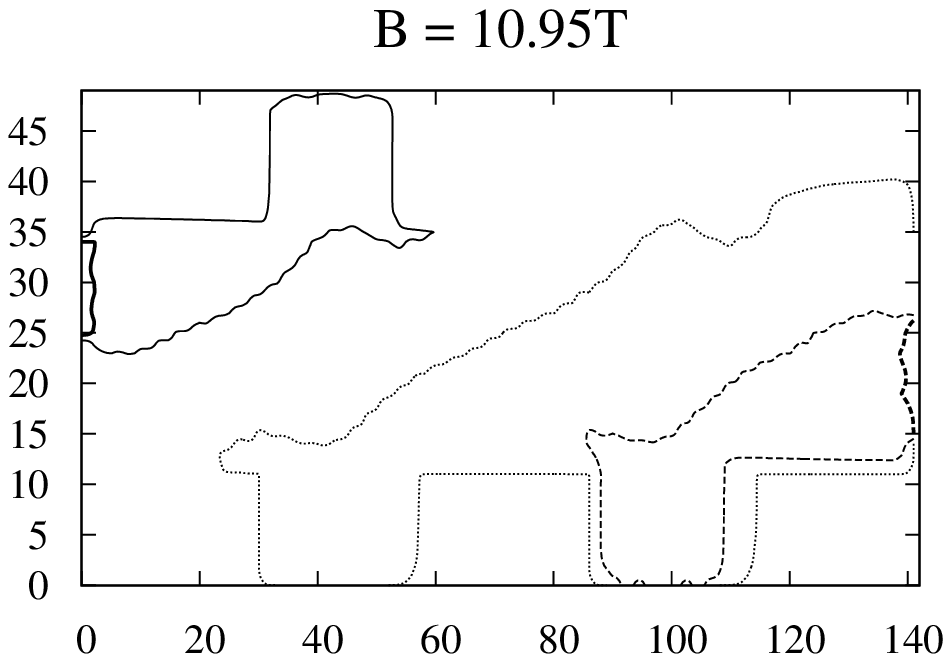} \\
  	 	\includegraphics[scale=0.7]{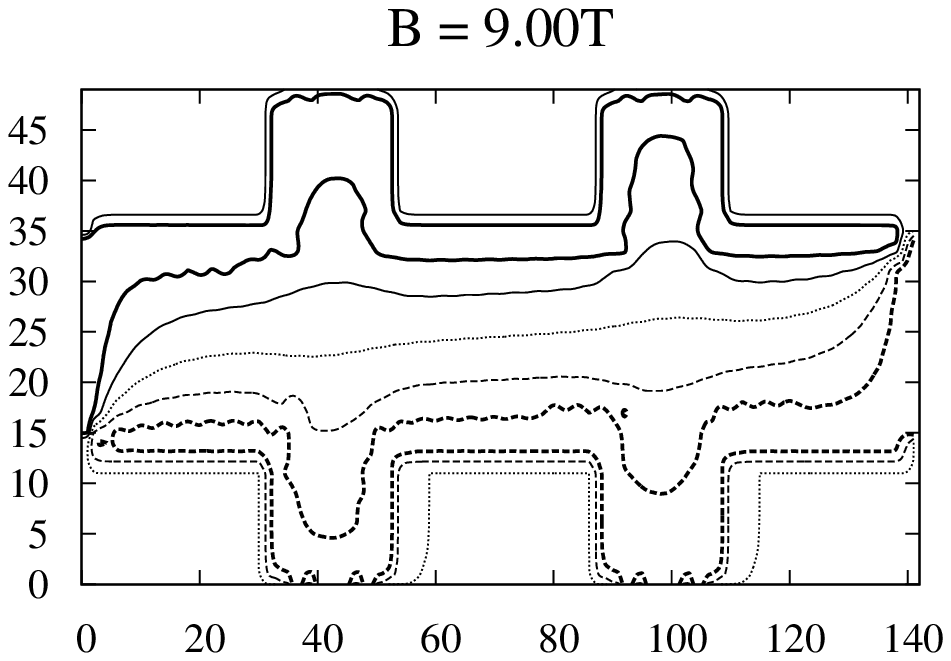} &
				\includegraphics[scale=0.7]{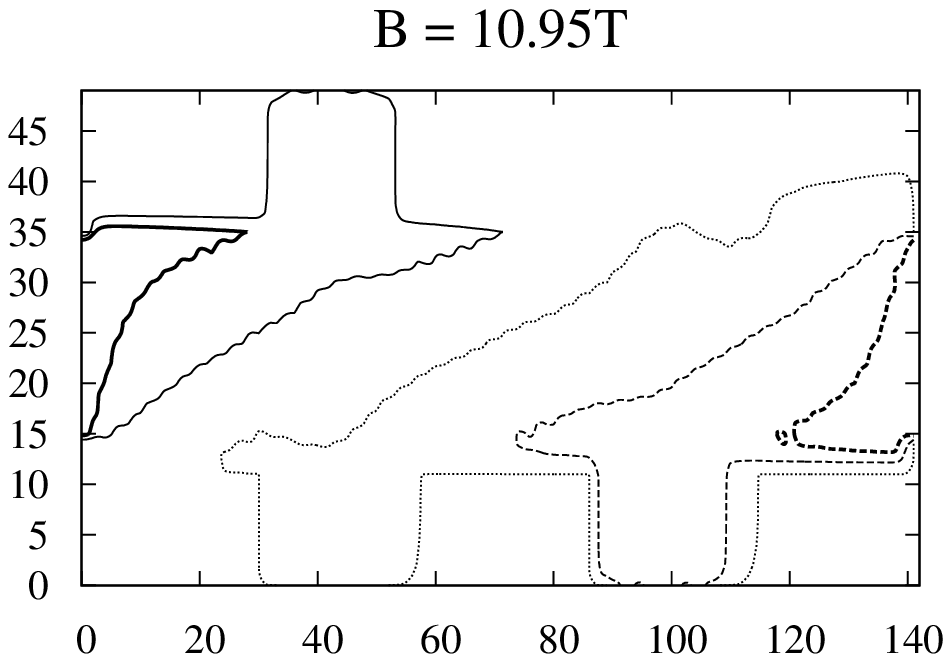}
   	\end{tabular}
   \caption{}
   \label{f_contacts}
 \end{center}
\end{figure}

\begin{figure}
 \begin{center}
   \includegraphics[scale=0.7]{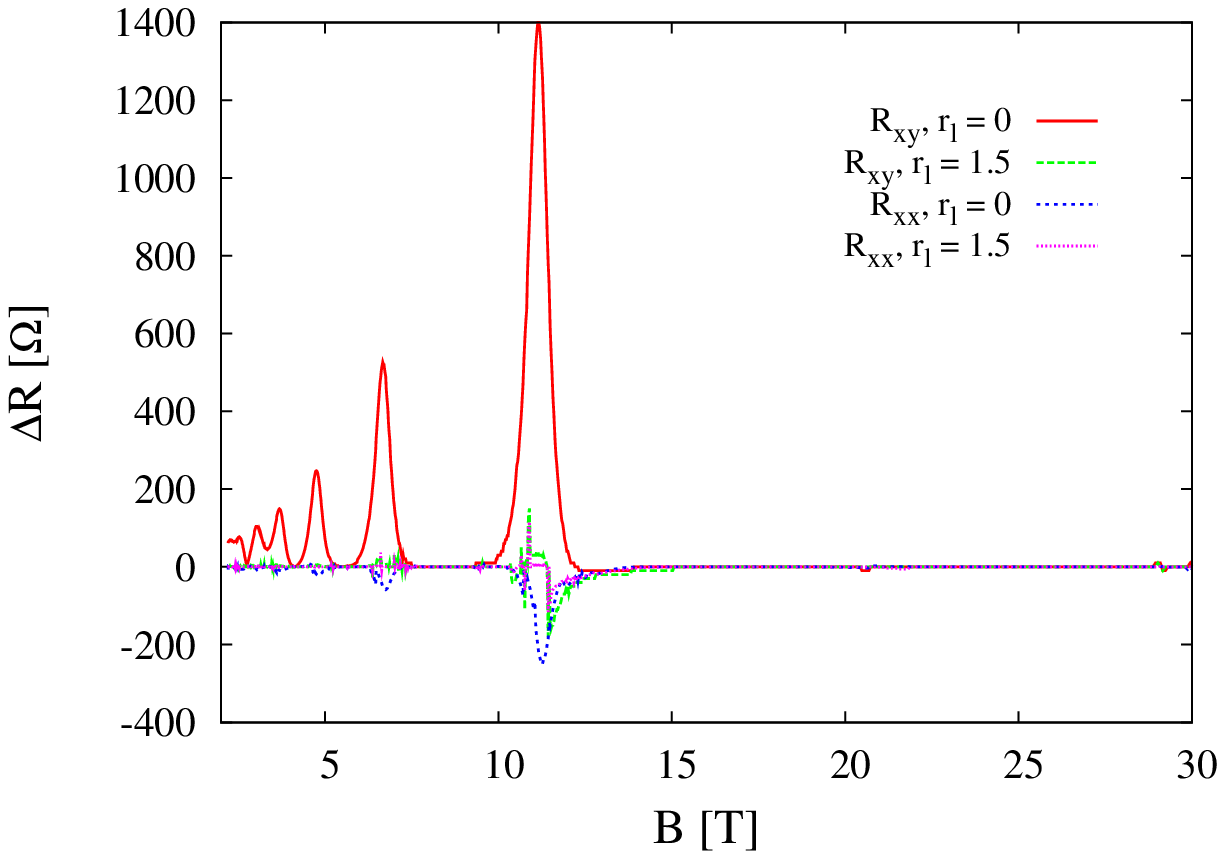}
   \caption{}
   \label{f_RContactdiff}
 \end{center}
\end{figure}

\begin{figure}
 \begin{center}
   \begin{tabular}{cc}   				   				
   		\includegraphics[scale=0.7]{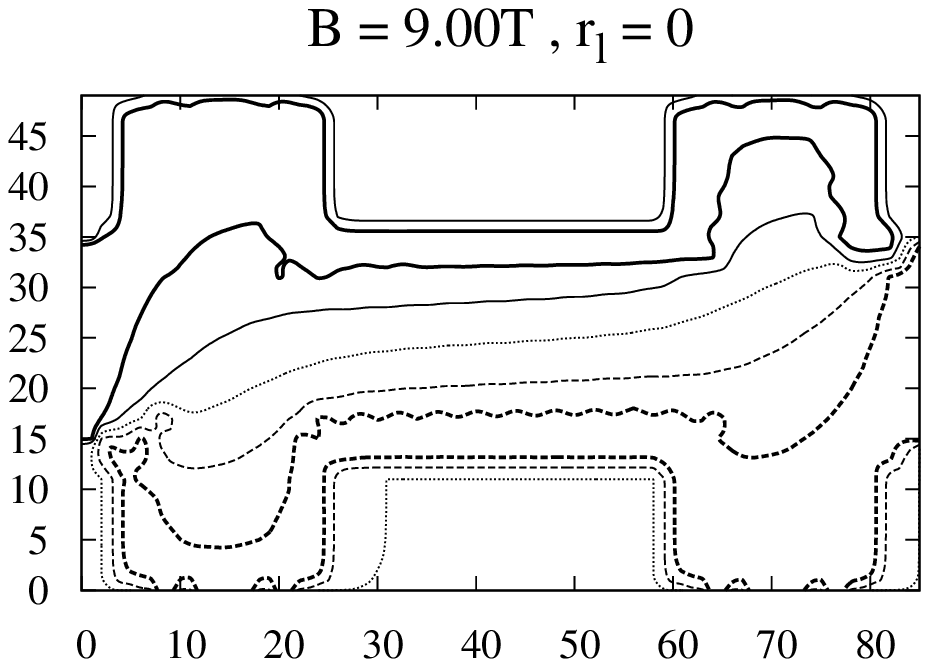} &
  			\includegraphics[scale=0.7]{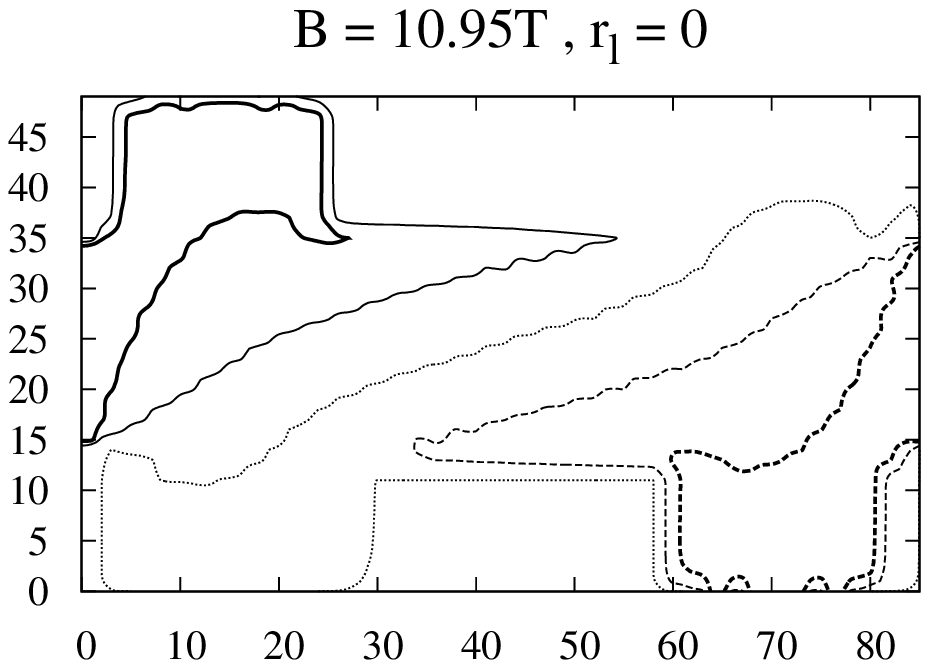} \\
  		\includegraphics[scale=0.7]{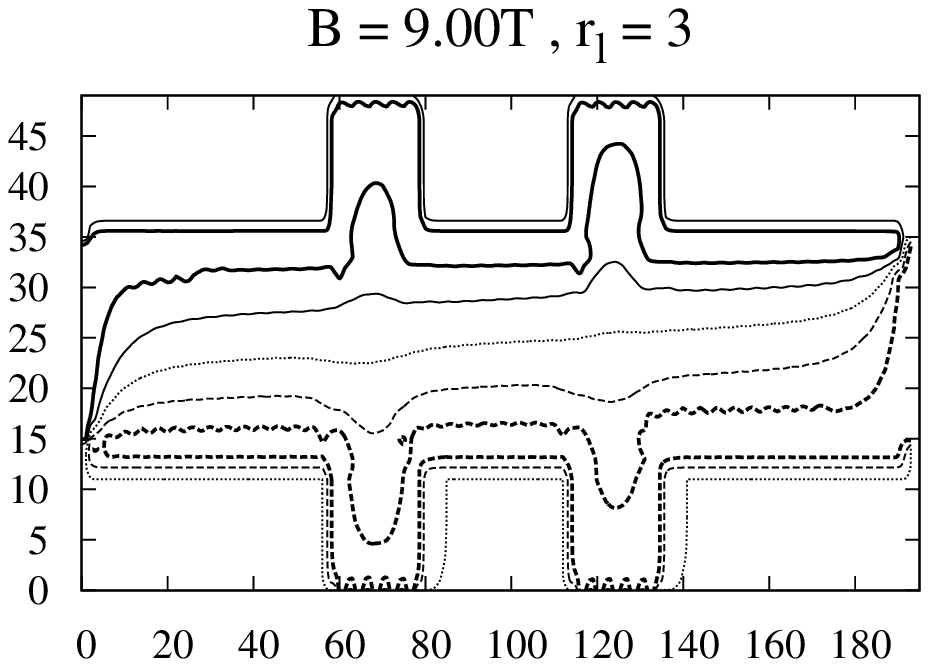} &
  	 		\includegraphics[scale=0.7]{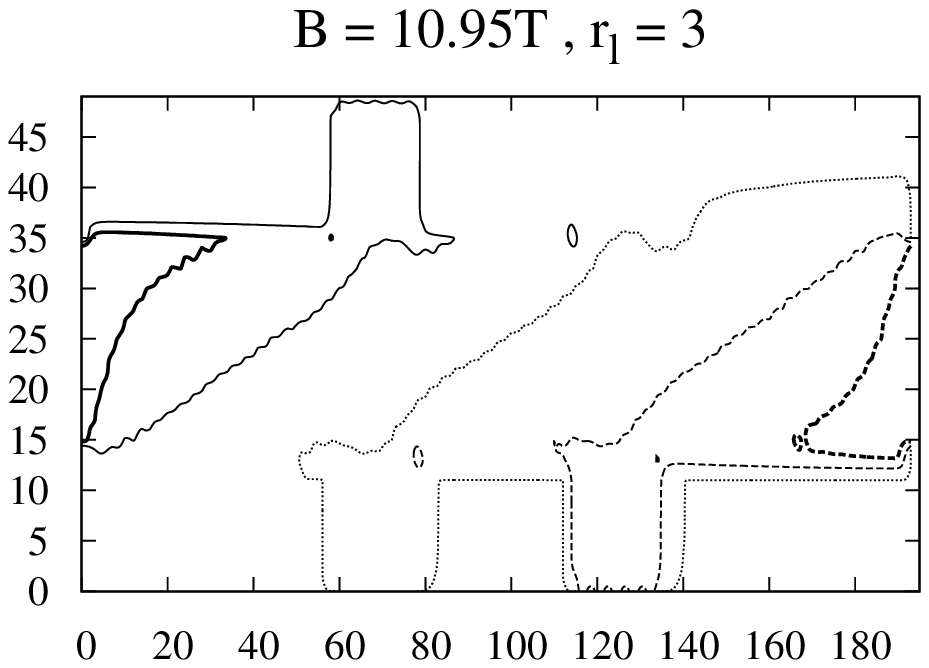}
   \end{tabular}
   \caption{}
   \label{f_lengthLead}
 \end{center}
\end{figure}

\begin{figure}
 \begin{center}
   \begin{tabular}{c}
   		\includegraphics[scale=0.7]{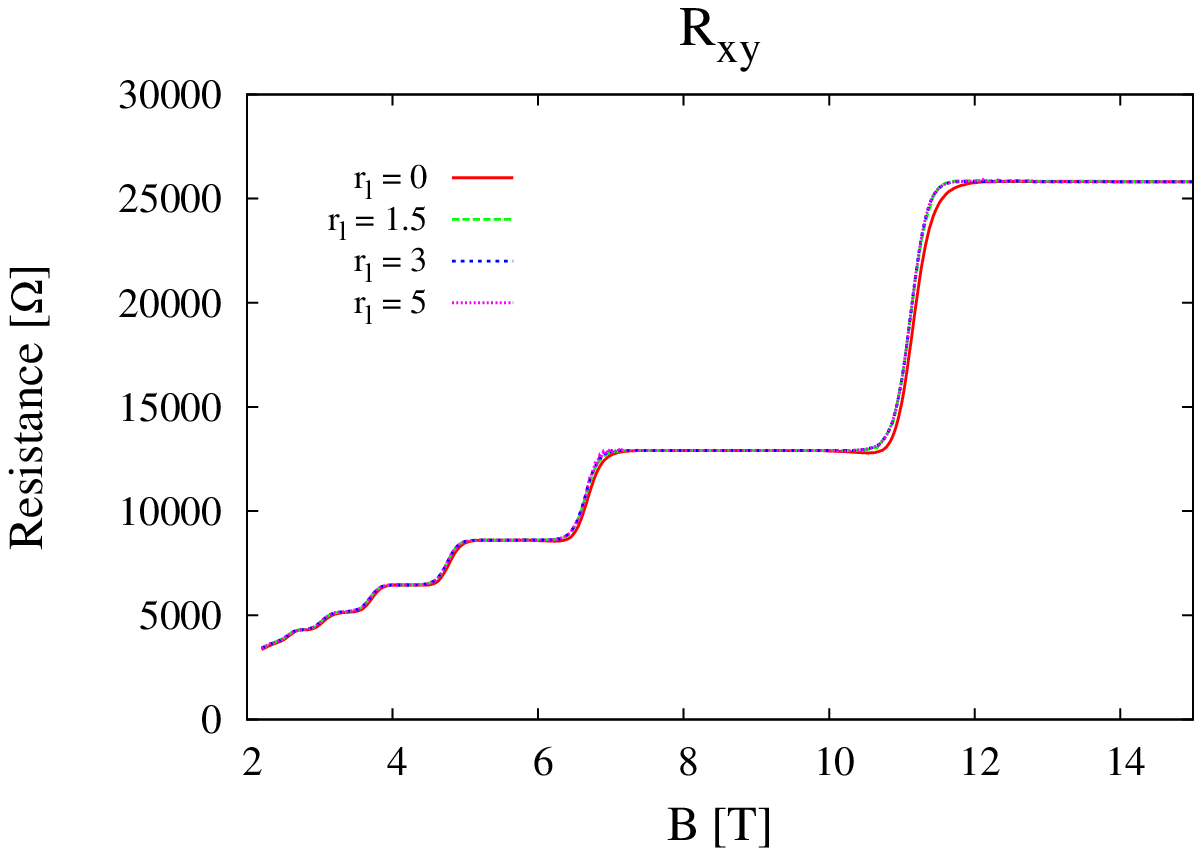} \\
   			\includegraphics[scale=0.7]{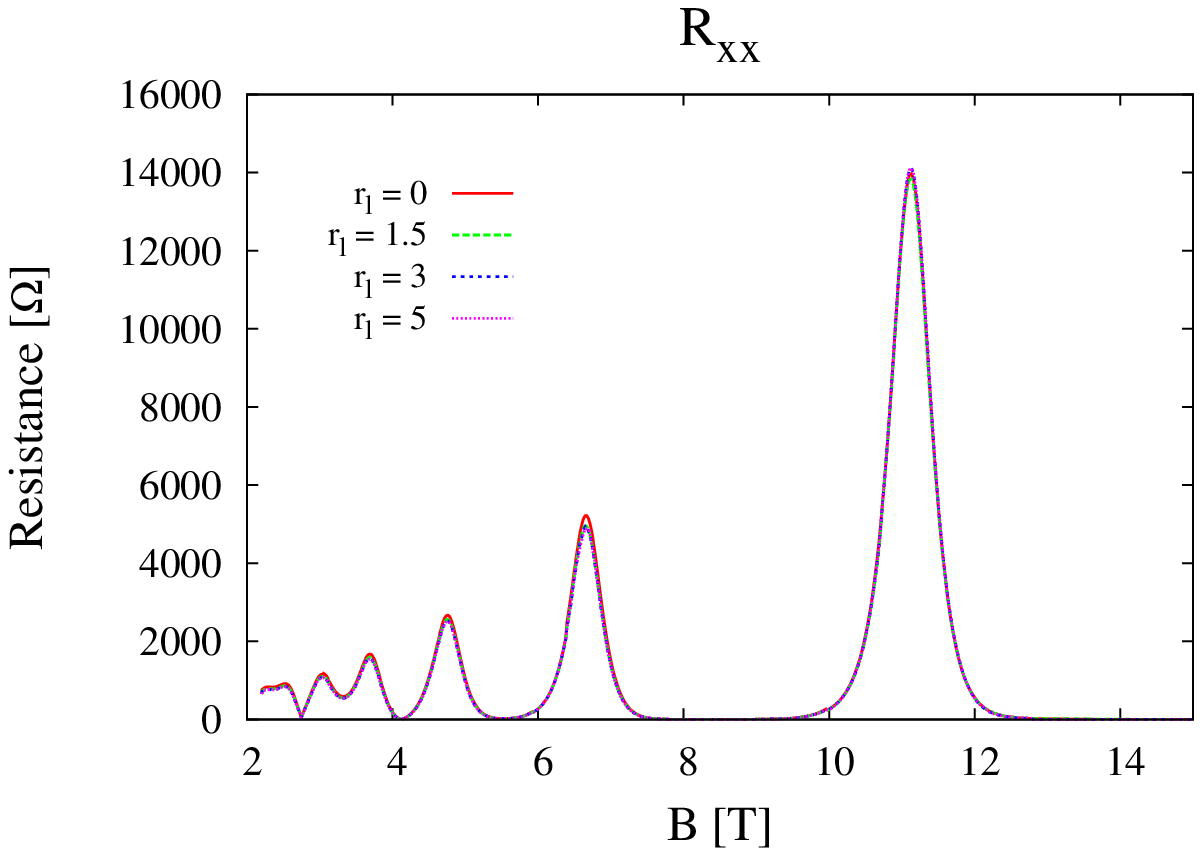}
   \end{tabular}   
   \caption{}
   \label{f_RLeads}
 \end{center}
\end{figure}

\begin{figure}
 \begin{center}
   \includegraphics[scale=0.7]{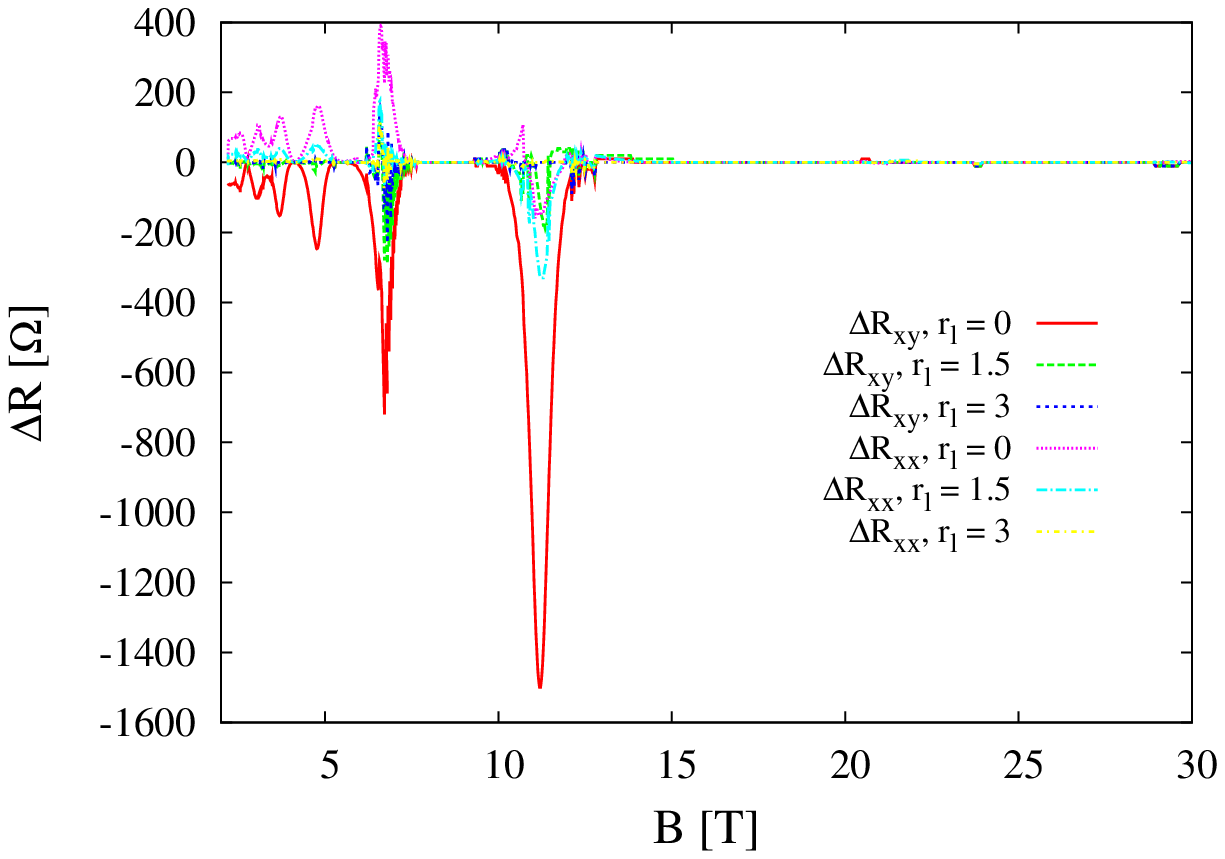}
   \caption{}
   \label{f_Rdiff}
 \end{center}
\end{figure}

\begin{figure}
 \begin{center}
   \begin{tabular}{cc}
   		\includegraphics[scale=0.7]{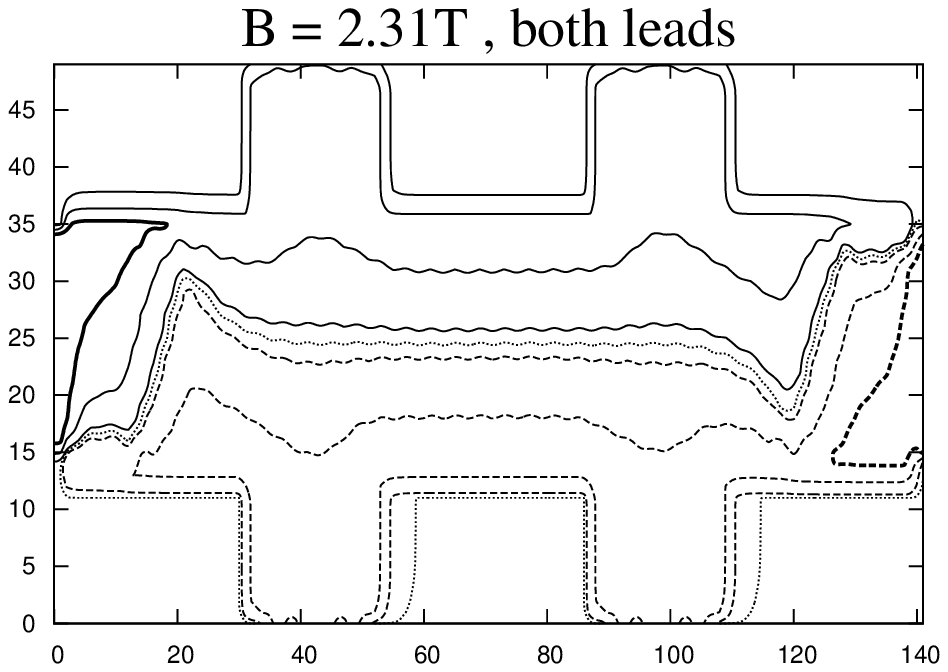} &
   			\includegraphics[scale=0.7]{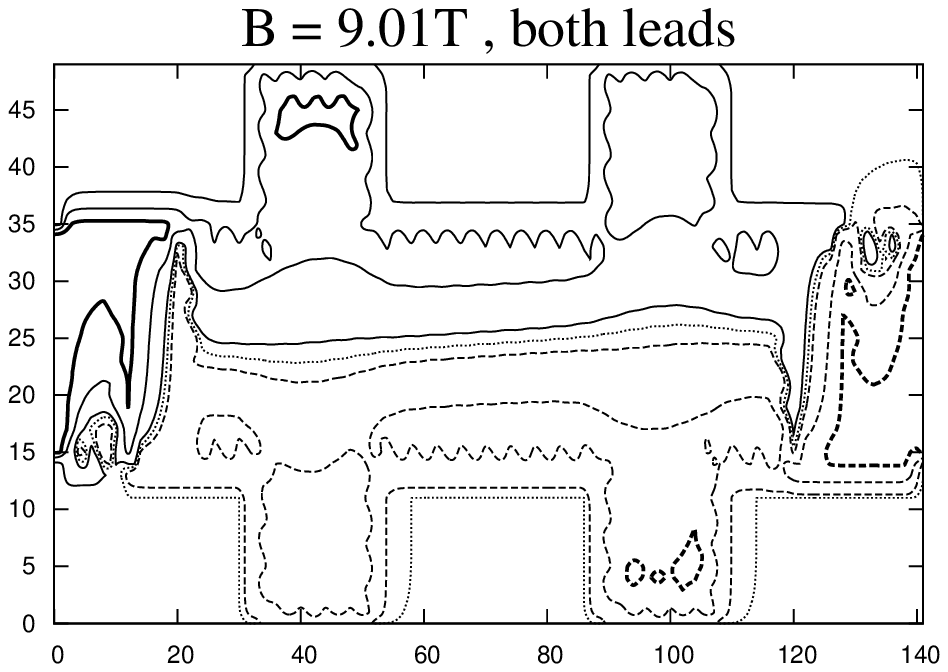} \\
  	 	\includegraphics[scale=0.7]{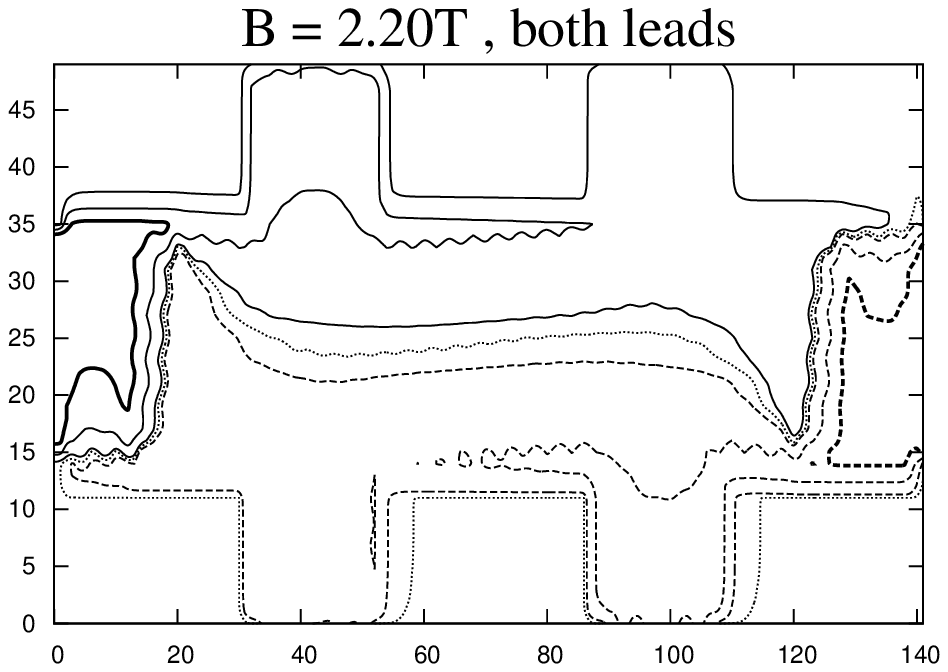} &
  	 		\includegraphics[scale=0.7]{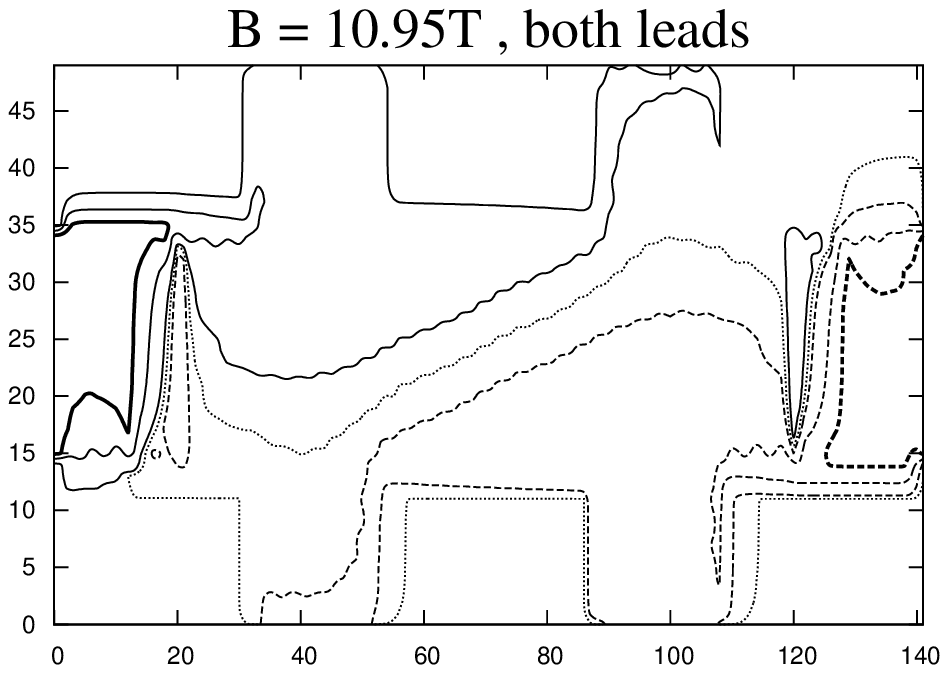} \\
  	 	\includegraphics[scale=0.7]{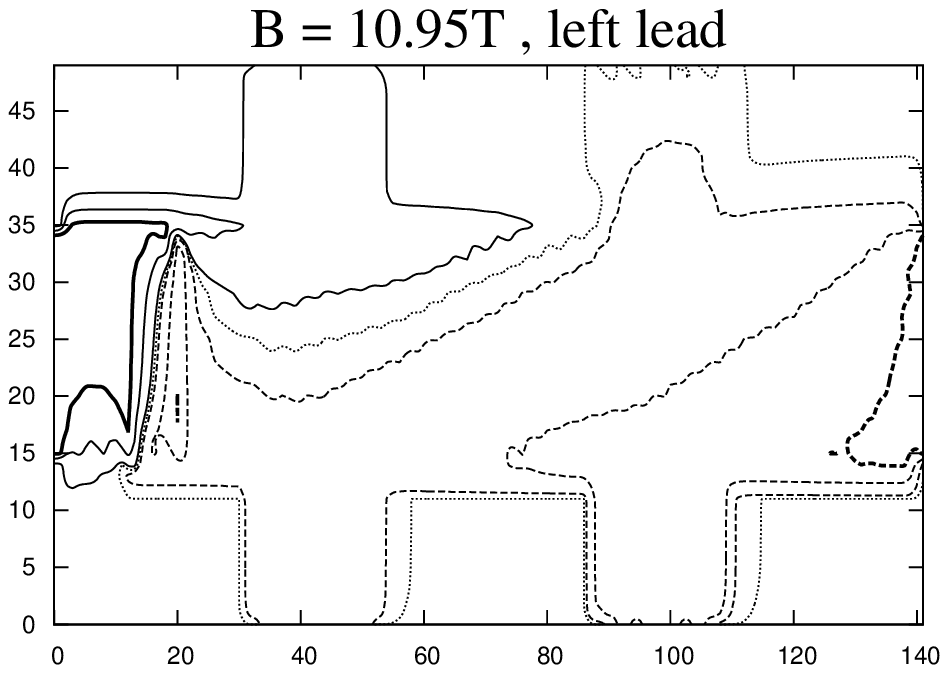} &
  	 		\includegraphics[scale=0.7]{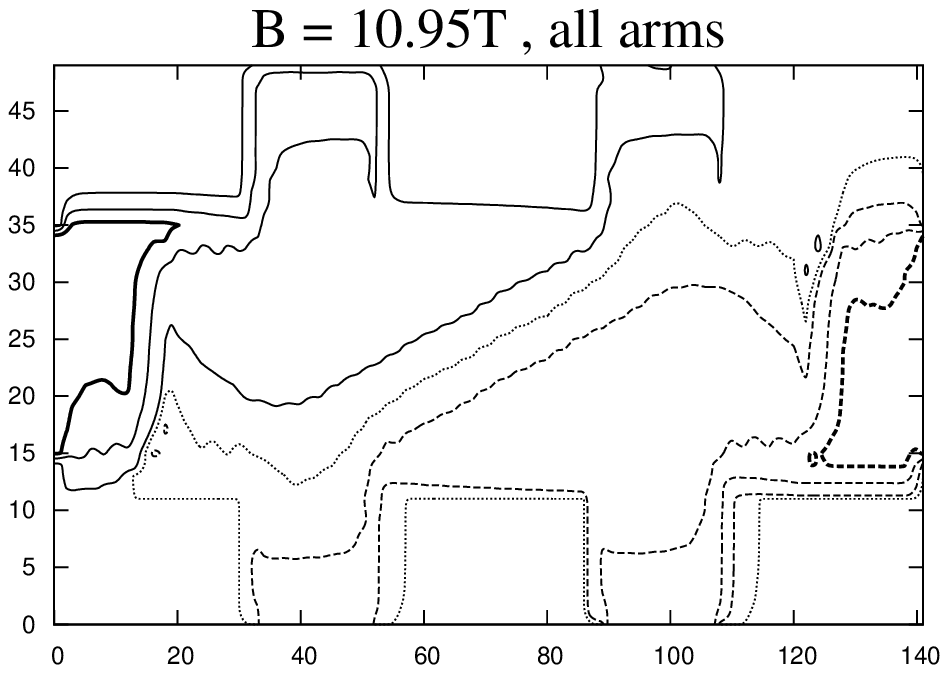}
   	\end{tabular}
   \caption{}
   \label{f_barrier}
 \end{center}
\end{figure}

\begin{figure}
 \begin{center}
   \includegraphics[scale=0.7]{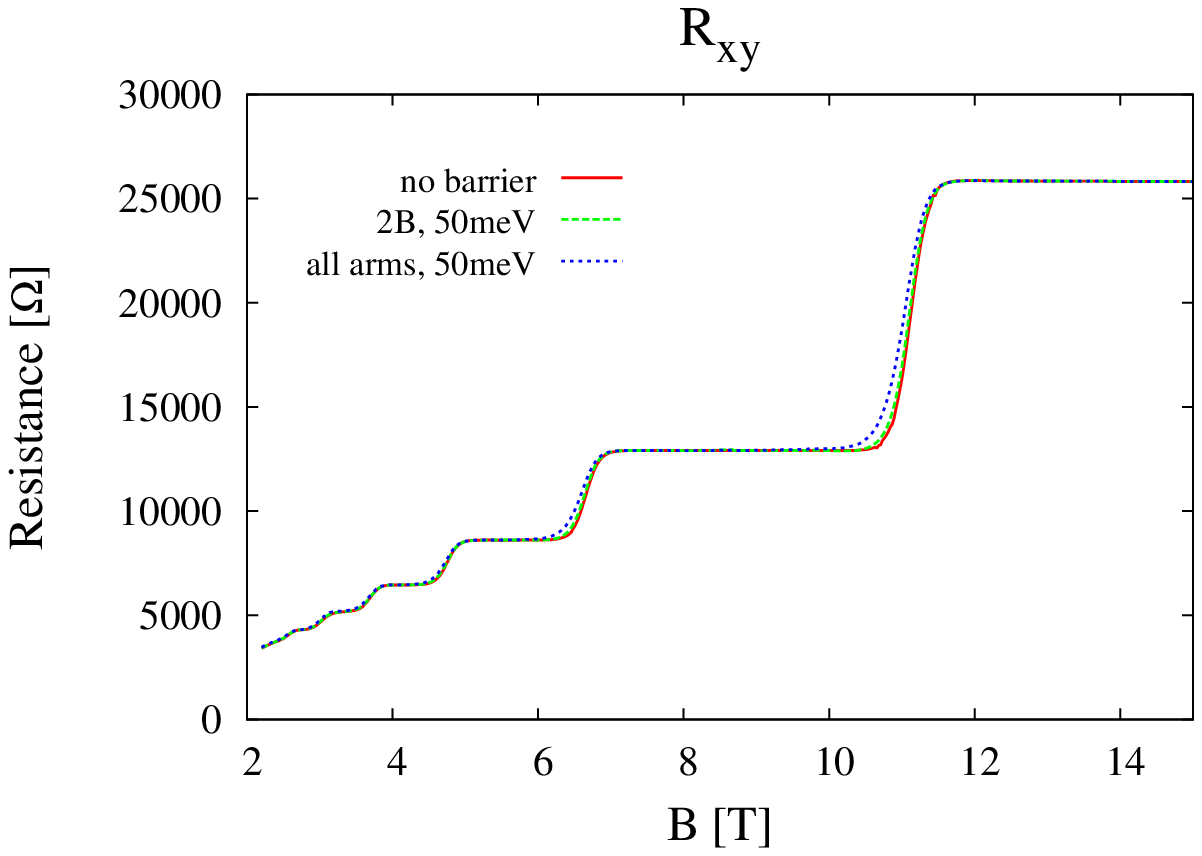}
   \caption{}
   \label{f_RBarriers}
 \end{center}
\end{figure}

\begin{figure}
 \begin{center}
   \includegraphics[scale=0.7]{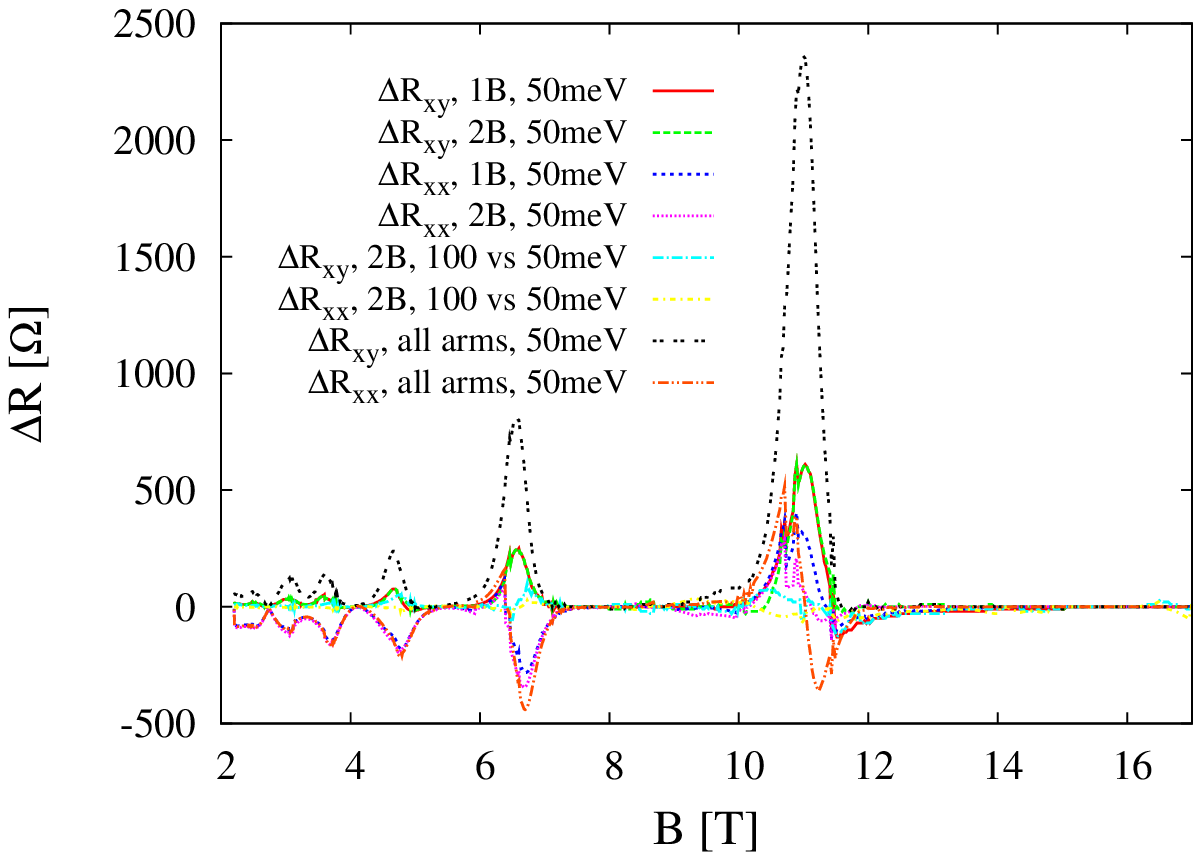}
   \caption{}
   \label{f_BarrierRdiff}
 \end{center}
\end{figure}

\begin{figure}
 \begin{center}
   \includegraphics[scale=0.7]{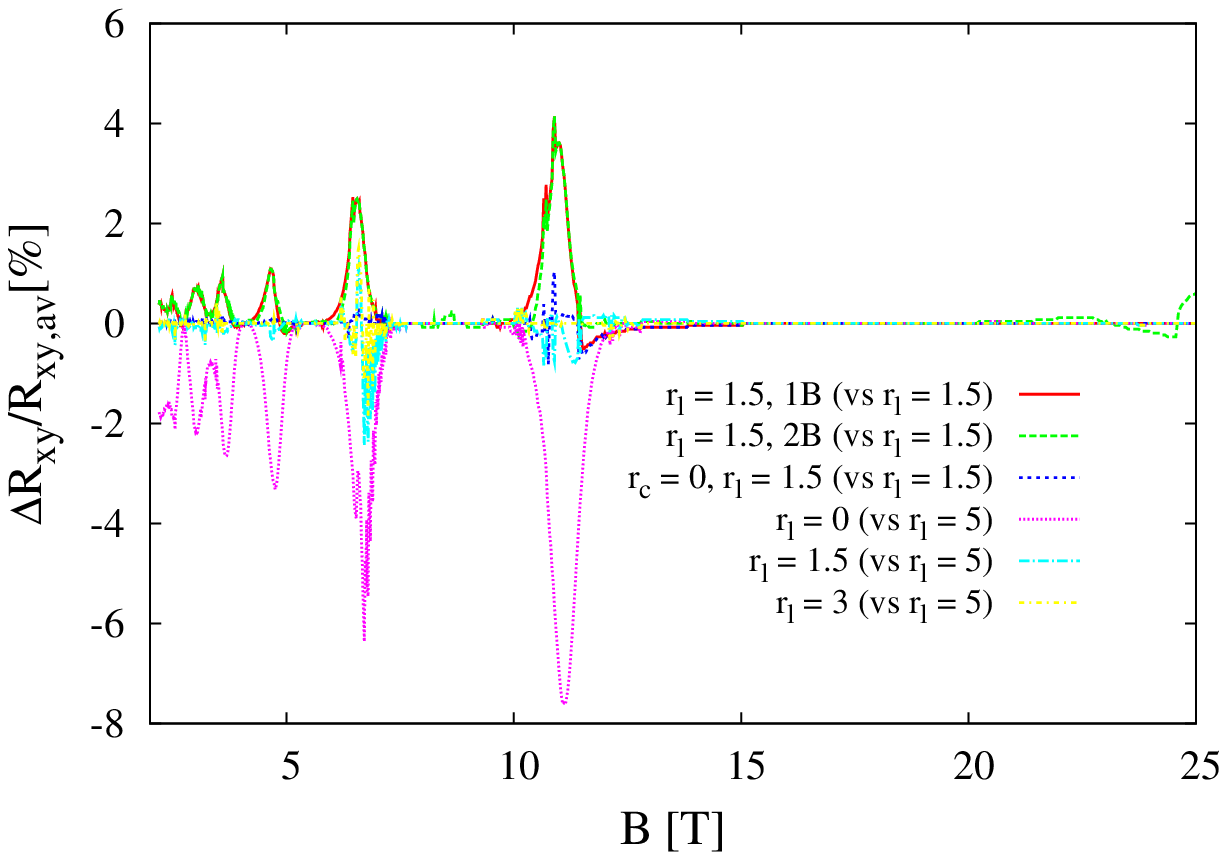}
   \caption{}
   \label{f_RrelErr}
 \end{center}
\end{figure}

Figure 4.(color online)
Visualization of chemical potential distributions across the sample with different current-injecting leads. The upper row shows results for the ratio length versus width of $r_l\approx 0$ whereas the lower row demonstrates results for $r_l=3$. The magnetic field is set to values of the plateu ($9T$) and transition ($11T$) region. Corresponding resistances as functions of magnetic field are shown in Fig.~\ref{f_RLeads}. The linestyle and values of contours is equivalent to Fig.~\ref{f_contacts}. 
 
Figure 5. (color online)
Longitudinal and transversal resistance versus magnetic field for various values $r_l$ of the length of the current injecting leads.

Figure 6. (color online)
Difference of transversal and longitudinal resistances for leads with various $r_l$ relative to values for $r_l=5$, plotted against magnetic field.

Figure 7. (color online)
Potential distributions for current-injecting contacts having a barrier in the lead. The first row shows the plateau region whereas the second demonstrates results for the transition region. The last row shows the case of only one barrier and the sample with barriers in the leads and measurement arms. The linestyle of contours is equivalent to Fig.~\ref{f_contacts}, however we used the values $0,\pm 1,\pm 5,\pm 9$.

Figure 8. (color online)  
Transversal resistance versus magnetic field for various configurations of barriers.

Figure 9. (color online)
Difference of transversal and longitudinal resistances for leads with and without barrier in one or both current-injecting leads, plotted against magnetic field. The energies denote the value of the height of the unscreened barriers.

Figure 10. (color online)
Relative Error of transversal resistances for various $r_l$, with and without barrier, and the case of leads with point contacts ($r_c=0$), no barrier and $r_l=1.5$, compared to leads with $r_c=1$, no barrier and $r_l=1.5$, plotted against magnetic field.

Figure 11. (color online) 
This figure shows the setup of gated sample 1. The contacts are numbered by starting with 1 at the left boundary edge and increasing the numbers in the clockwise direction.
The external potentials are connected via contacts 1 and 2 and the (longitudinal) voltage drop is measured between contacts 4 and 3. Gating potentials are indicated by shaded stripes
along contacts.

\begin{figure}
 \begin{center}
   \includegraphics[scale=0.5]{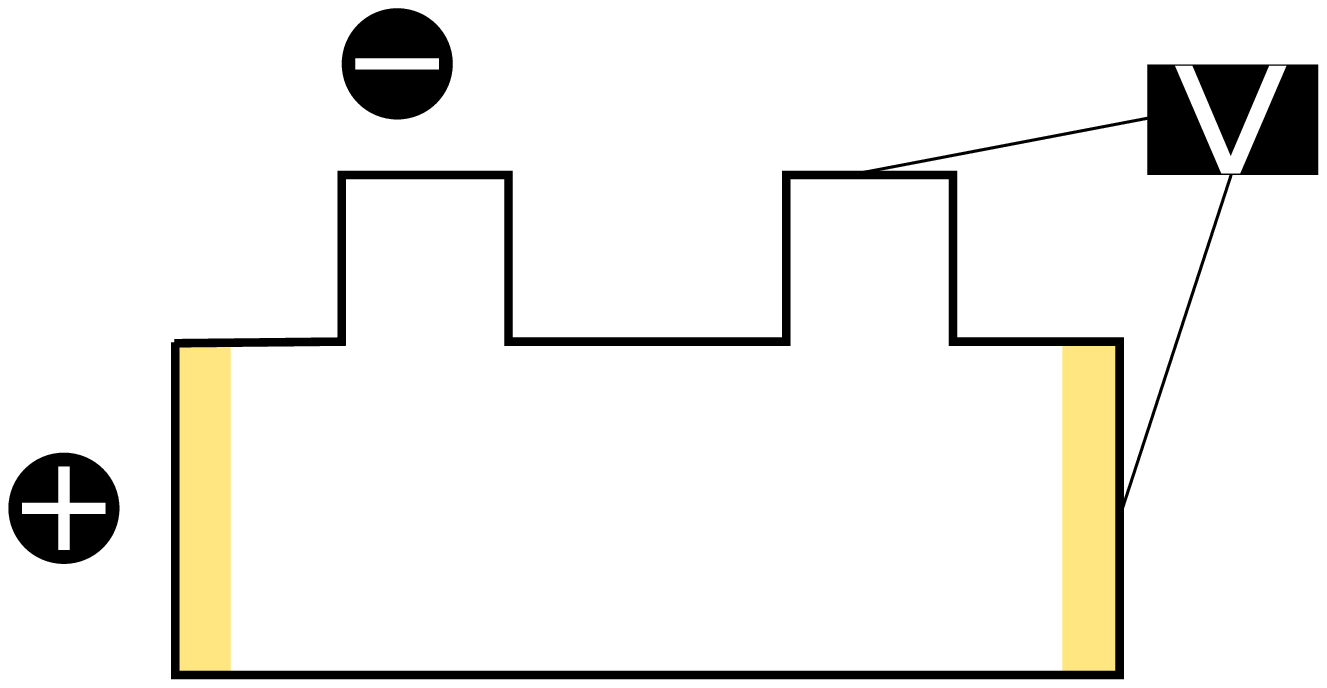}
   \caption{}
   \label{f_sample1}
 \end{center}
\end{figure}

Figure 12. (color online) 
This figure shows the setup of gated sample 2 for various gating configurations. The contacts are numbered as in Fig.~\ref{f_sample1}. The external potentials are connected via contacts 2 and 4 and the (Hall) voltage drop is measured between contacts 1 and 3. Gating potentials are indicated by shaded stripes
along contacts. Gating configurations are labelled as a) contacts 1,4, b) contacts 2,3 and c) all contacts.

Figure 13. (color online)
The longitudinal resistance $R_L$ as a function of the applies magnetic field, calculated from sample 1 by dividing the voltage drop by the total current. We compare the ideal sample (without gatings) with the gated configuration in Fig.~\ref{f_sample1}. 

Figure 14. (color online)
Visualization of chemical potential distribution across sample 1 as a color chart. The color label gives the respective voltage values. The magnetic field is indicated above each plot. 

\begin{figure}
 \begin{center}
   \begin{tabular}{cc}
   	\includegraphics[scale=0.5]{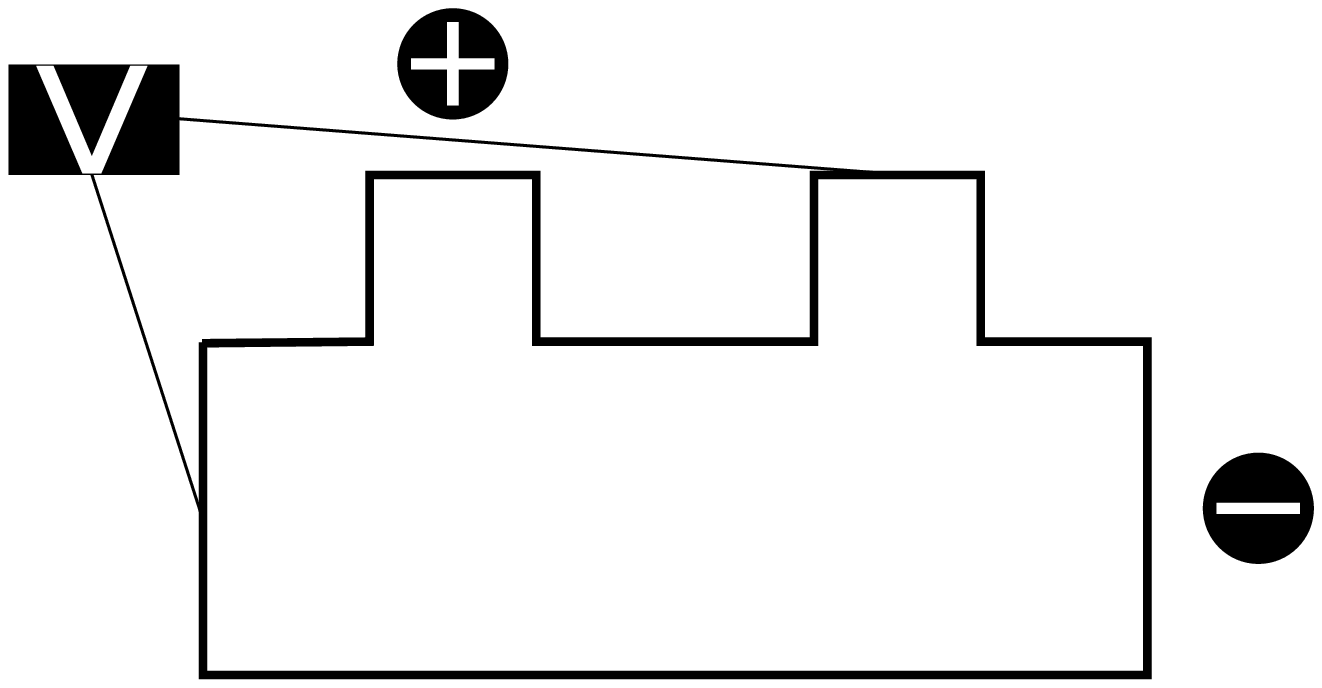} &
   		\includegraphics[scale=0.5]{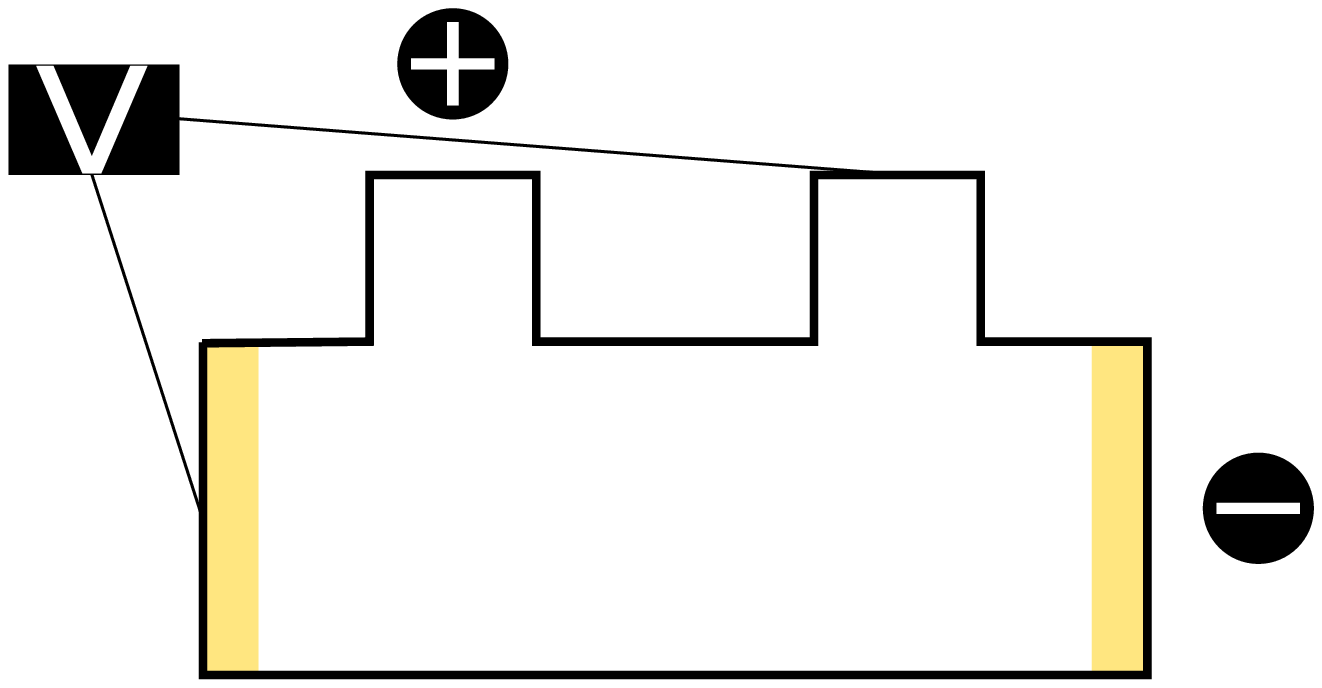} \\
   	\includegraphics[scale=0.5]{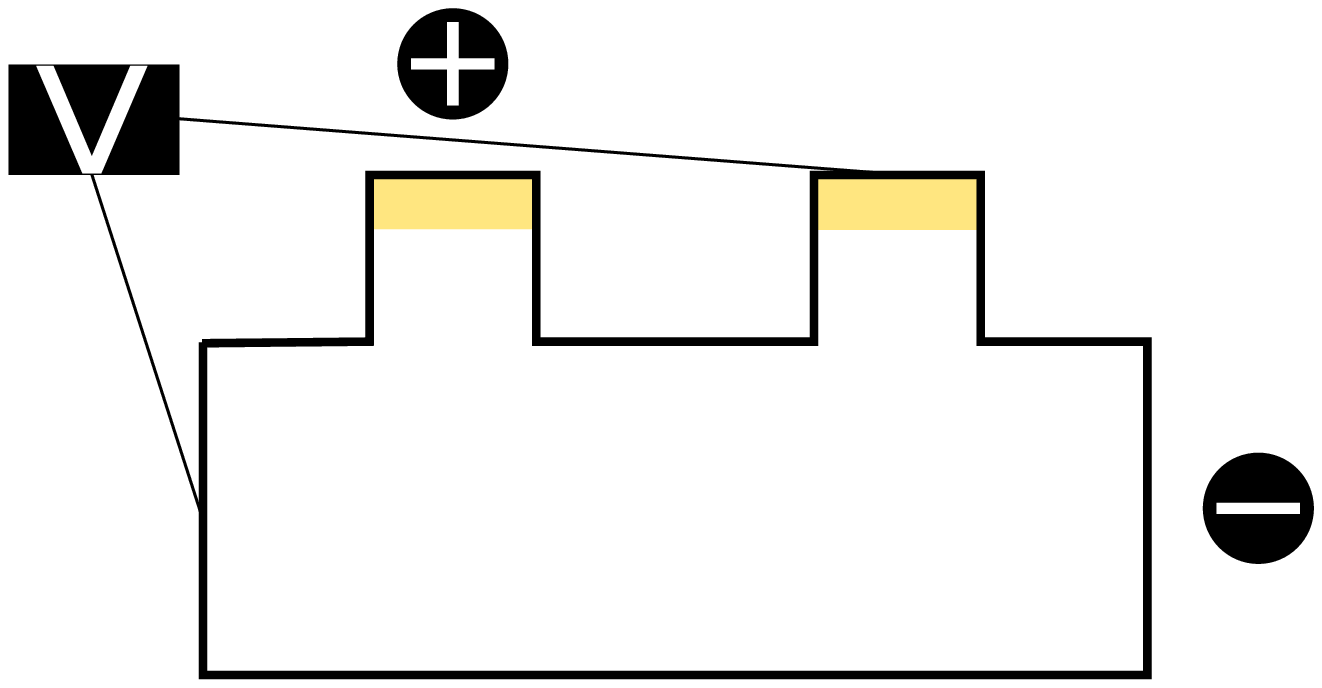} &
   		\includegraphics[scale=0.5]{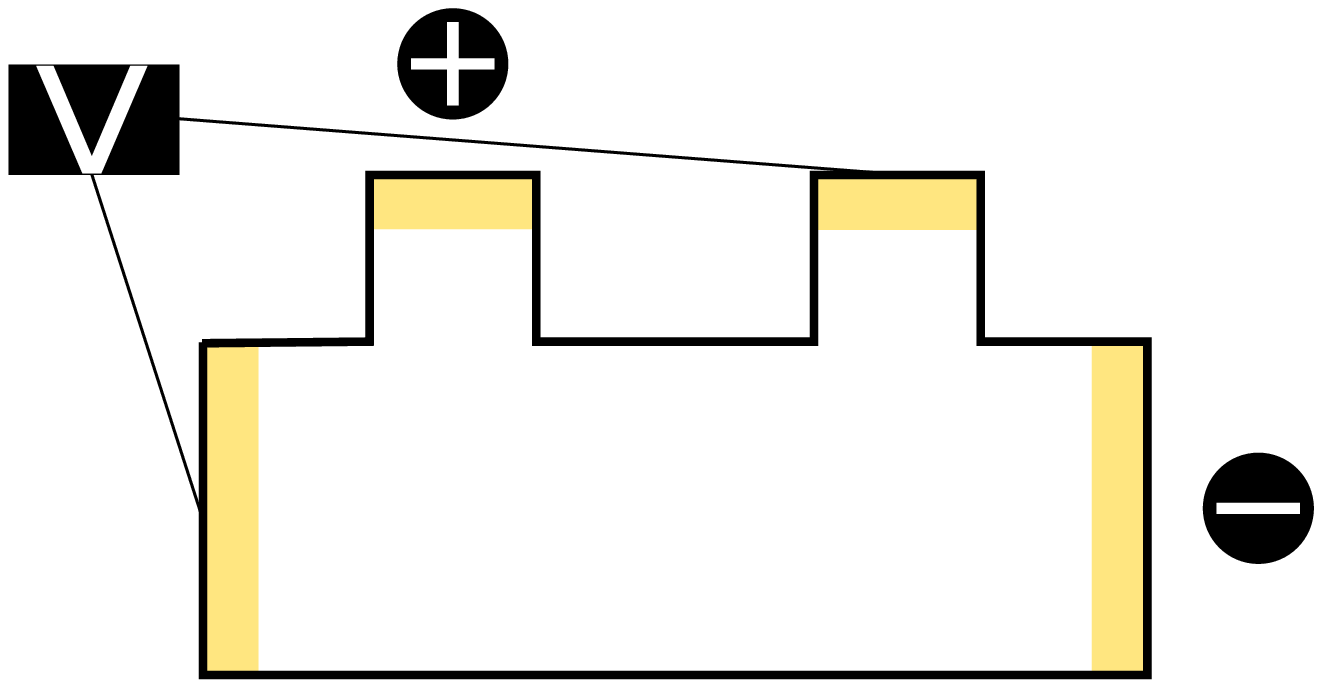}
   \end{tabular}
   \caption{}
   \label{f_sample2}
 \end{center}
\end{figure}

\begin{figure}
 \begin{center}
   \includegraphics[scale=1]{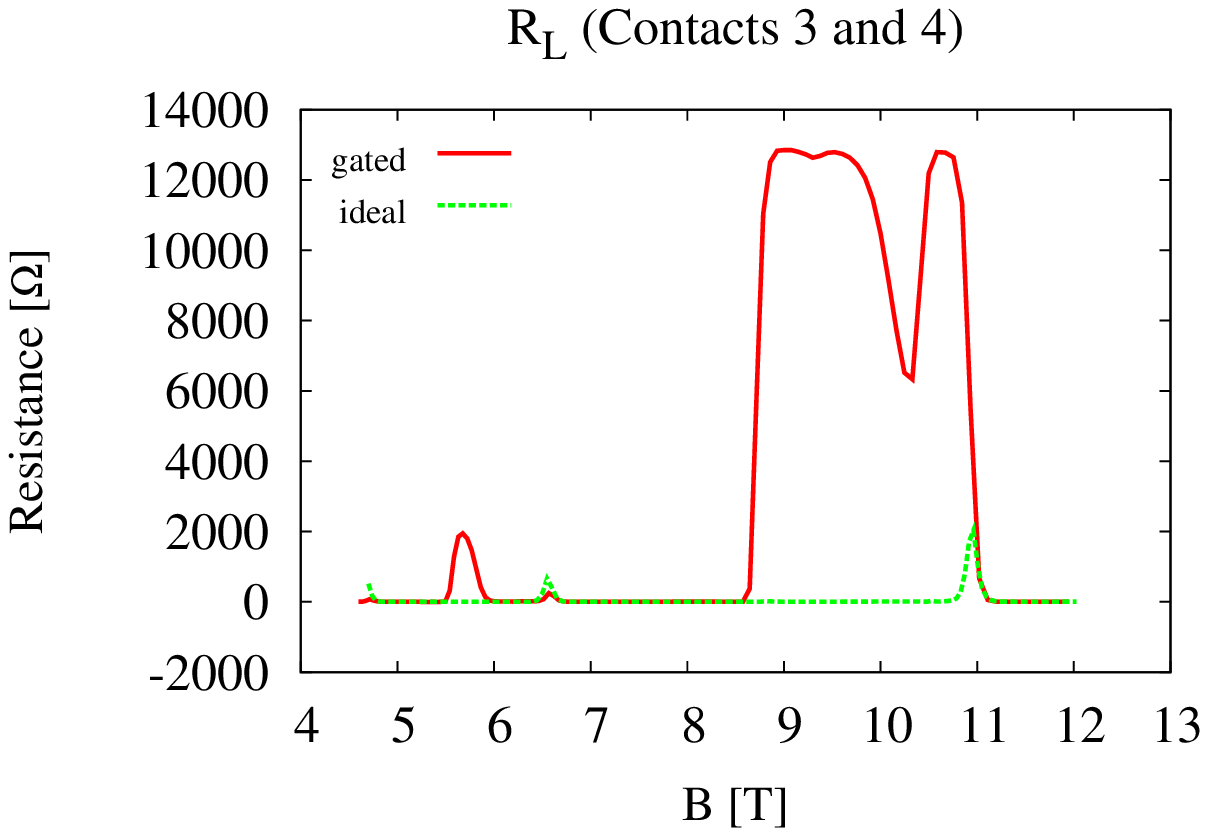}
   \caption{}
   \label{f_Rxx}
 \end{center}
\end{figure}

\begin{figure}
 	\begin{center}
   	\begin{tabular}{cc}
   		\includegraphics[scale=0.7]{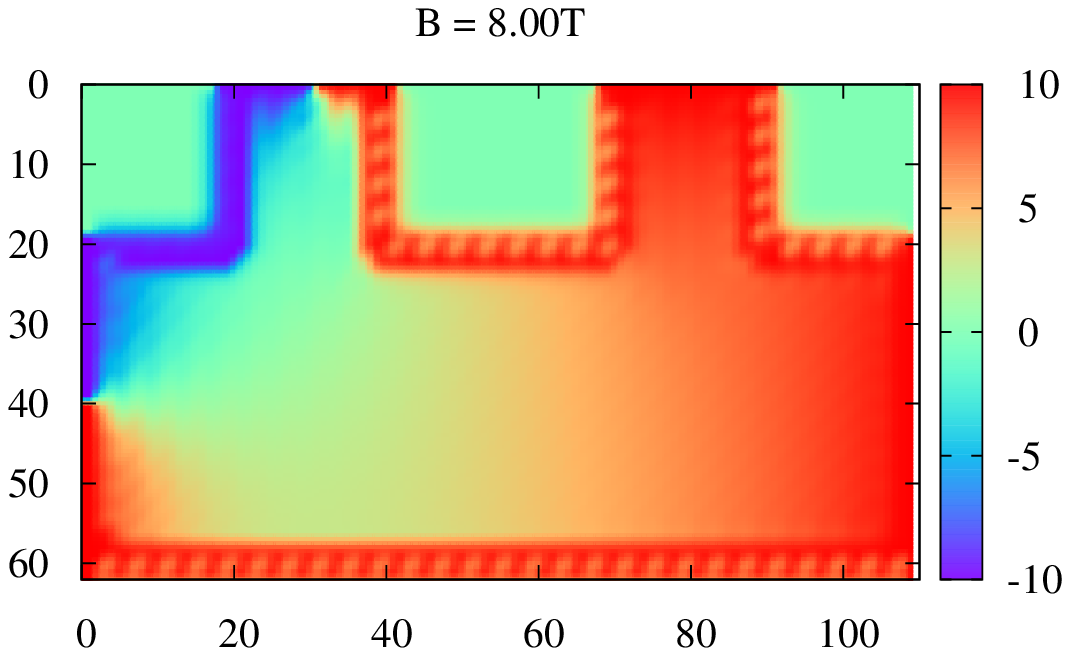} &
   			\includegraphics[scale=0.7]{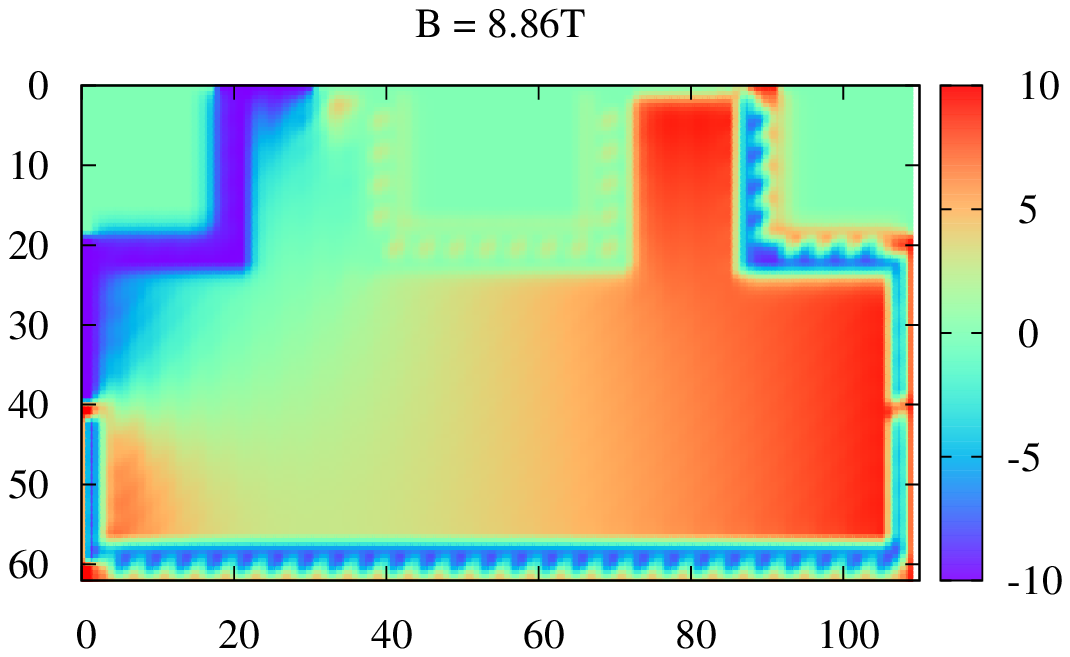} \\
   			\includegraphics[scale=0.7]{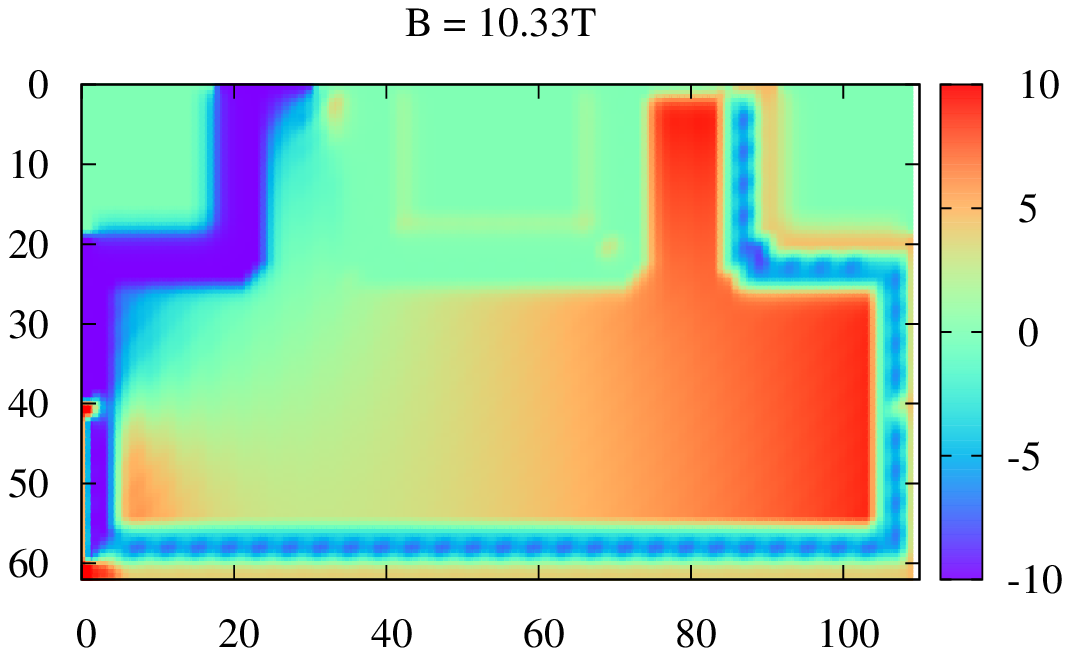} &
   			\includegraphics[scale=0.7]{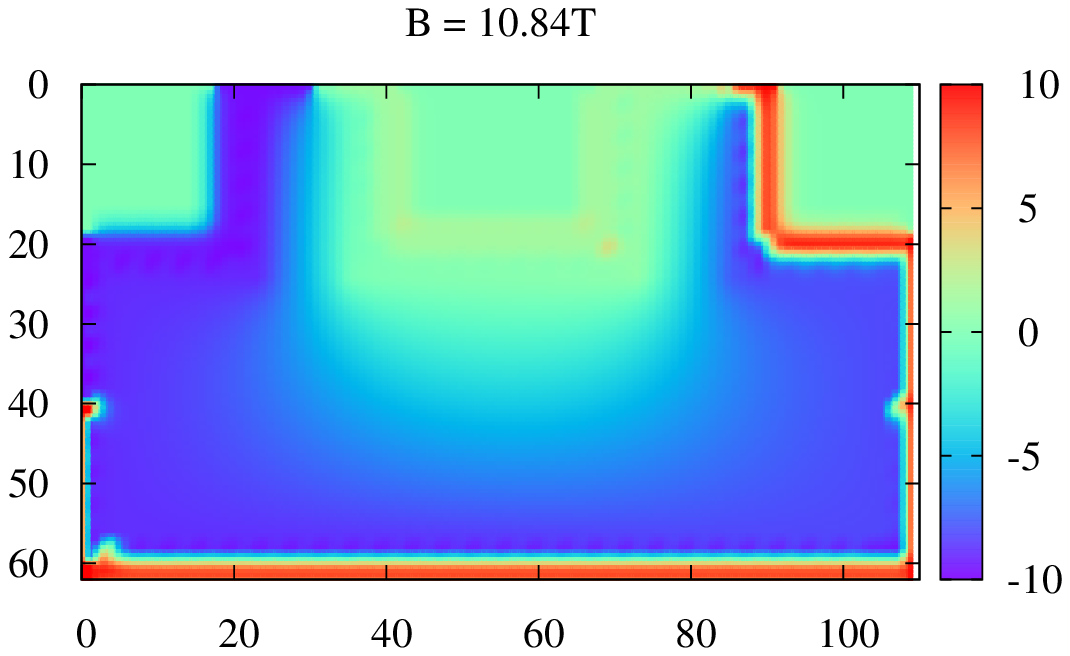}
   	\end{tabular}
   	\caption{}
   	\label{f_PotDistL}
 	\end{center}
\end{figure}

\begin{figure}
 \begin{center}
   \includegraphics[scale=1]{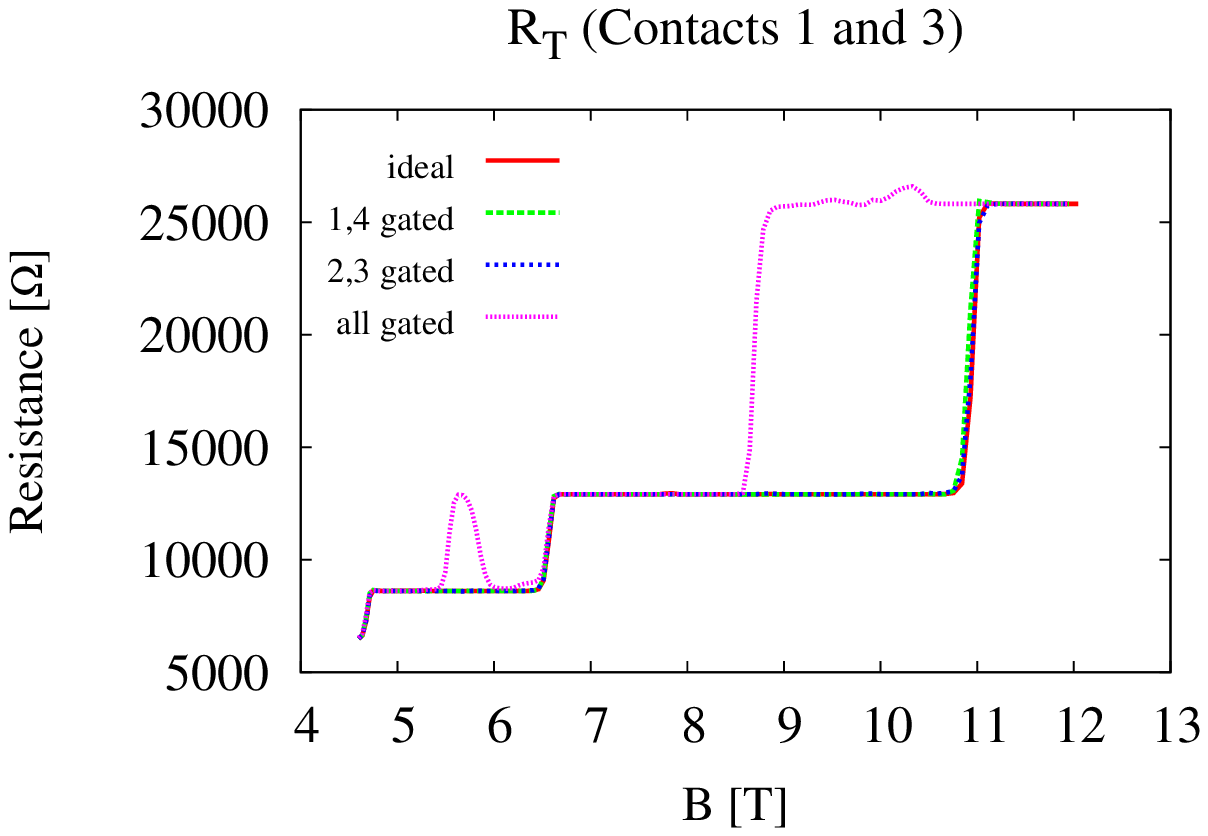}
   \caption{}
   \label{f_Rxy}
 \end{center}
\end{figure}

\begin{figure}
 \begin{center}
   \begin{tabular}{cc}
			\includegraphics[scale=0.7]{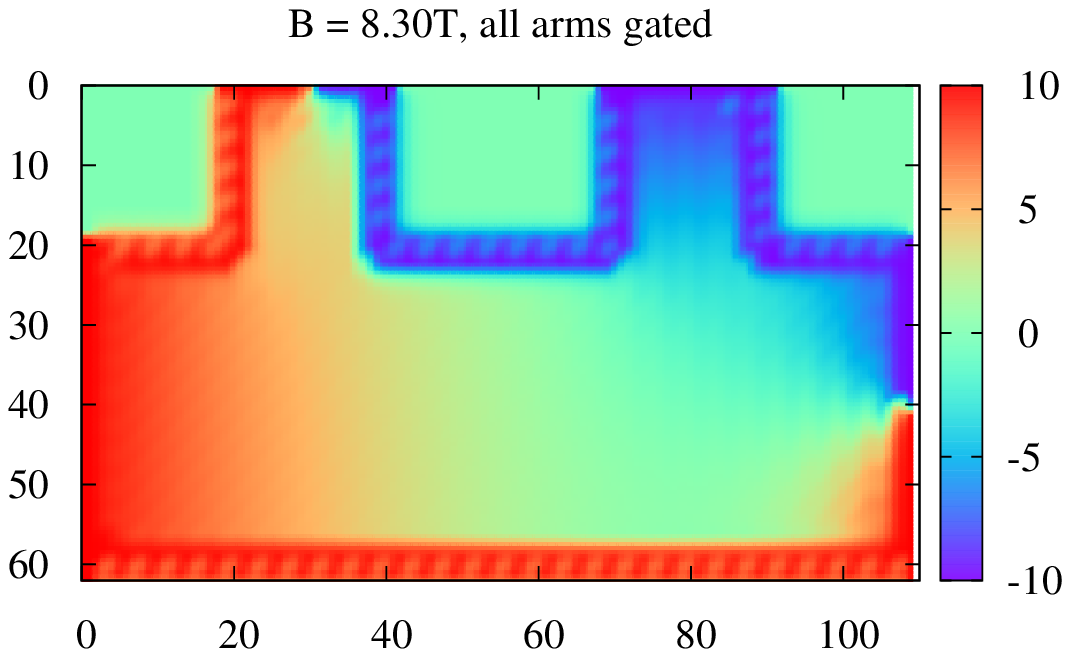} &
   			\includegraphics[scale=0.7]{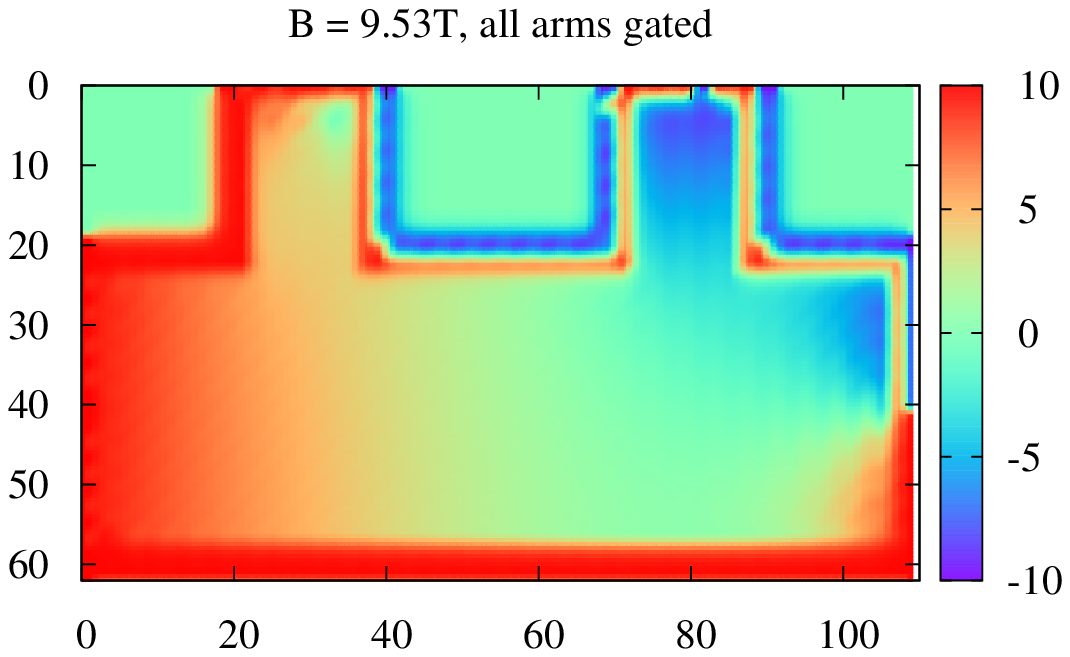} \\
   		\includegraphics[scale=0.7]{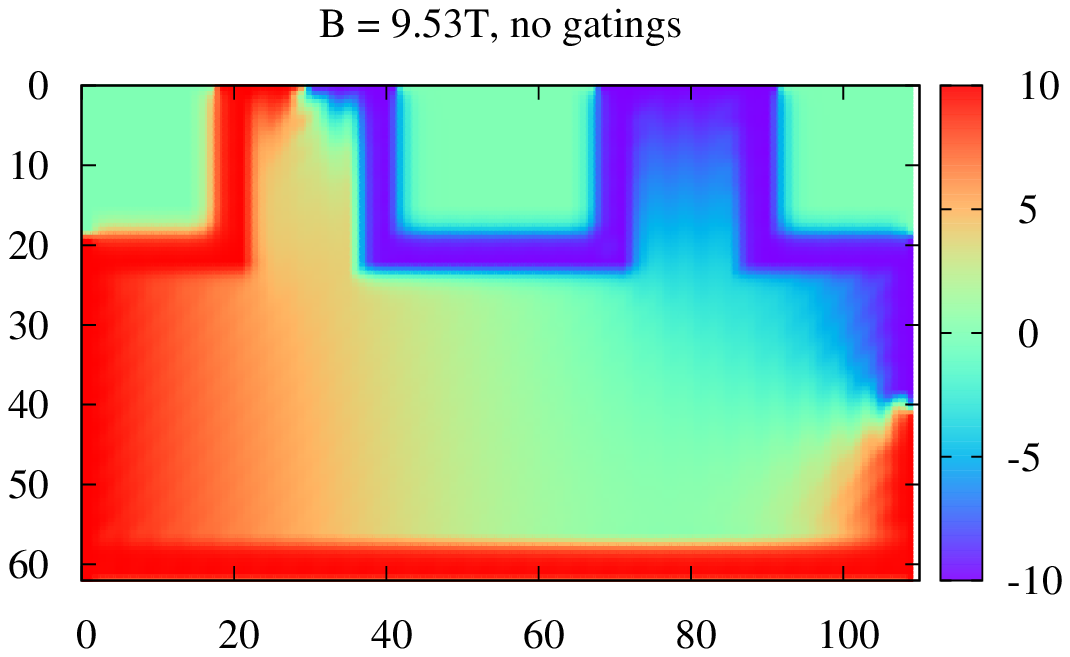}
   	\end{tabular}
   \caption{}
   \label{f_decoupledEC}
 \end{center}
\end{figure}

Figure 15. (color online)
The Hall resistance $R_T$ as a function of the applies magnetic field, calculated from sample 2 by dividing the voltage drop by the total current. We compare various possible configurations of gatings with the numbers corresponding to Fig.~\ref{f_sample2} and ideal corresponds to no gating at all. 

Figure 16. (color online)
Visualization of chemical potential distribution across the sample as a color chart. The color label gives the respective voltage values. The upper row corresponds to all contacts gated while the remaining image corresponds to no gating. Magnetic fields are indicated above each plot.
\newpage  




\end{document}